# On the monotonicity, log-concavity and tight bounds of the generalized Marcum and Nuttall $Q-$functions

Yin Sun*, *Student Member, IEEE,* Árpád Baricz, and Shidong Zhou*, *Member, IEEE.*

*Abstract*—In this paper, we present a comprehensive study of the monotonicity and log-concavity of the generalized Marcum and Nuttall $Q-$functions. More precisely, a simple probabilistic method is firstly given to prove the monotonicity of these two functions. Then, the log-concavity of the generalized Marcum $Q-$function and its deformations is established with respect to each of the three parameters. Since the Nuttall $Q-$function has similar probabilistic interpretations as the generalized Marcum $Q-$function, we deduce the log-concavity of the Nuttall $Q-$function. By exploiting the log-concavity of these two functions, we propose new tight lower and upper bounds for the generalized Marcum and Nuttall $Q-$functions. Our proposed bounds are much tighter than the existing bounds in the literature in most of the cases. The relative errors of our proposed bounds converge to 0 as $b \to \infty$. The numerical results show that the absolute relative errors of the proposed bounds are less than 5% in most of the cases. The proposed bounds can be effectively applied to the outage probability analysis of interference-limited systems such as cognitive radio and wireless sensor network, in the study of error performance of various wireless communication systems operating over fading channels and extracting the log-likelihood ratio for differential phase-shift keying (DPSK) signals.

## I. INTRODUCTION

THE generalized Marcum $Q-$function has a long history in the study of target detection probability in radar communications [1], [2]. Recently, it has gained much attention for its important applications in digital communications over fading channels, such as the bit error performance analysis dealing with differential coherent and non-coherent detection [3]-[11], the efficient performance evaluation of dual selective combining diversity over correlated Rayleigh and Nakagami-$m$ fading channels [8] and [12]-[13], the information-theoretic analysis of Rician MIMO channels [14]-[15], the energy detection of unknown signals over fading channels [16] and the primary user detection performance for cognitive radio with multiple antennas [17]. For $a \geq 0$ and $b \geq 0$ the generalized Marcum $Q-$function of real order $\nu > 0$ is defined by [18]

$$Q_\nu(a,b) = \frac{1}{a^{\nu-1}} \int_b^\infty t^\nu e^{-\frac{t^2+a^2}{2}} I_{\nu-1}(at) dt, \quad (1)$$

where $I_\nu$ is the modified Bessel function of the first kind of order $\nu$ and the right hand side of the above equation is replaced by its limiting value when $a = 0$. If $\nu = 1$, then this reduces to the standard (first-order) Marcum $Q-$function, denoted as $Q(a,b)$. The standard Nuttall $Q-$function is a generalization of the Marcum $Q-$function, and is defined by [19, eq. 86]

$$Q_{\mu,\nu}(a,b) = \int_b^\infty t^\mu e^{-\frac{t^2+a^2}{2}} I_\nu(at) dt, \quad (2)$$

where $b, \mu, \nu \geq 0$ and $a > 0$. A normalized version of the Nuttall $Q-$function with respect to the parameter $a$ is defined as [20], [21]

$$\mathcal{Q}_{\mu,\nu}(a,b) = \frac{Q_{\mu,\nu}(a,b)}{a^\nu} = \int_b^\infty \frac{t^\mu}{a^\nu} e^{-\frac{t^2+a^2}{2}} I_\nu(at) dt. \quad (3)$$

In particular, if $\mu = \nu + 1$, then $\mathcal{Q}_{\mu,\nu}(a,b)$ reduces to the generalized Marcum $Q-$function of order $\nu + 1$, i.e. we have $\mathcal{Q}_{\nu+1,\nu}(a,b) = Q_{\nu+1}(a,b)$ for all admissible values of $a, b$ and $\nu$. The applications for the standard and normalized Nuttall $Q-$function include array processing performance in fading channels with co-channel interference [22], the largest eigenvalue of noncentral complex Wishart matrices [23], performance analysis of multiple-input-multiple-output (MIMO) systems employing multichannel beamforming in arbitrary-rank Ricean channels [24] and polynomial approximations for evaluation of the average error probability over slow-fading channels [25]. Some more applications of the Nuttall $Q-$function can be found in [21] and in the references therein.

However, the precise computations of the generalized Marcum and Nuttall $Q-$functions are quite difficult, mainly because the modified Bessel function of the first kind $I_\nu$ is involved in the integrands of (1) and (2). In the last few decades, many researchers were working on precise and stable numerical calculation algorithms of the generalized Marcum and Nuttall $Q-$functions (see [26]-[31] for example). While only integer order generalized Marcum and Nuttall $Q-$functions are considered in these papers, appropriate evaluating methods for non-integer order generalized Marcum and Nuttall $Q-$functions are desirable for the performance analysis

Manuscript received October 21, 2008; revised October 22, 2009.
* Corresponding authors.
Y. Sun and S. Zhou's research was supported by National Natural Science Foundation of China (60832008), National Basic Research Program of China (2007CB310608) and Program for New Century Excellent Talents in University (NCET). Á. Baricz's research was supported by the János Bolyai Research Scholarship of the Hungarian Academy of Sciences.
Y. Sun and S. Zhou are with the State Key Laboratory on Microwave and Digital Communications, Tsinghua National Laboratory for Information Science and Technology, and Department of Electronic Engineering, Tsinghua University, Beijing, China. Address: Room 4-407, FIT Building, Tsinghua University, Beijing 100084, People's Republic of China. e-mail: sunyin02@mails.tsinghua.edu.cn, zhousd@tsinghua.edu.cn.
Á. Baricz is with the Department of Economics, Babeş-Bolyai University, Cluj-Napoca 400591, Romania. e-mail: bariczocsi@yahoo.com.
The material of this paper was presented in part at the IEEE Global Communications Conference 2008, New Orleans, LA, U.S.A., November 30th-December 4th, 2008 (IEEE GLOBECOM 2008).
Communicated by G. Taricco, Associate Editor for Communications.

of some wireless communication systems. Some examples can be found in [12]-[13], [21] and the references therein. In [32], an exact expression of $Q_\nu(a,b)$ was given for the case when $\nu$ is an odd multiple of 0.5. More compact closed-form expressions were proposed in [33] and [21]. A closed-form expression of $Q_{\mu,\nu}(a,b)$, where $\mu, \nu$ are odd multiples of 0.5 and $\mu \geq \nu$, was also given in [21]. During the review period of this paper, we discovered the papers [12] and [13], in which an finite-integral representation of $Q_\nu(a,b)$ with real-order $\nu$ was provided. But this finite-integral representation is still not in closed-form.

To obtain simpler evaluation methods for the generalized Marcum and Nuttall $Q-$functions, one is often willing to accept closed-form bounds of the functions if they are tight [34]. In recent years, many new lower and upper bounds were proposed as simpler alternative evaluating methods for the generalized Marcum and Nuttall $Q-$functions [21], [32]-[42]. Among them, many tight bounds for the Marcum $Q-$function and the generalized Marcum $Q-$function were obtained by exploiting the bounds of the integrand of (1) or modifying the integral region via a geometric interpretation of the functions [34]-[42]. However, these bounds can be only tight for a part of the region of $b$, i.e. either $b < a$ or $b > a$.

Recently, the monotonicity of the generalized Marcum and Nuttall $Q-$functions has drawn special interests to provide new bounds [21], [32]. Different analytical proofs of the monotonicity of the function $\nu \mapsto Q_\nu(a,b)$ were given in [21], [43]. However, this result was deduced also with a simple probabilistic method in [44], where it is also shown that the monotonicity of $\nu \mapsto Q_\nu(a,b)$ is equivalent to the monotonicity of the incomplete gamma function ratio, proved by Tricomi in 1950 [45]. A summary of the monotonicity of $Q_\nu(a,b)$ with respect to the parameters $\nu$, $a$ and $b$ can be found in [46, p. 451]. The monotonicity of $Q_{\mu,\nu}(a,b)$ on $\mu + \nu$, under the requirements that $a \geq 1$ and $\mu \geq \nu + 1$ and for constant difference $\mu - \nu$, was given in [21]. In this paper, a simple probabilistic method is provided which proves in an unified way the monotonicity of the generalized Marcum and Nuttall $Q-$functions with respect to different parameters. By this method, all the aforementioned results can be obtained, as well as a novel result that the normalized and standard Nuttall $Q-$functions are strictly increasing in $a$.

The bounds derived via the monotonicity of the generalized Marcum and Nuttall $Q-$functions hold true for all range of $b$. But they are not tight enough in terms of relative errors for a part of parameter region. Particularly for the case of large $b$, the relative errors of these bounds are unbounded (for the upper bounds) as $b$ approaches infinity.

In order to get even tighter bounds of the generalized Marcum and Nuttall $Q-$functions, some stronger property of these two functions needs to be established. In [4, p. 81], an asymptotic formula (but not a bound) for $Q_\nu(a,b)$ is provided when $b$ tends to infinity, given by

$$Q_\nu(a,b) \sim \left(\frac{b}{a}\right)^{\nu-0.5} Q(b-a), \qquad (4)$$

where $Q(\cdot)$ is the Gaussian $Q-$function and $\sim$ means that these two functions tend to be equal as $b$ increases.

This implies that $\sqrt{Q_{\nu-0.5}(a,b)Q_{\nu+0.5}(a,b)}$ can be used to estimate $Q_\nu(a,b)$ for integer $\nu$ and very large $b$, and the relative error of this estimation converges to 0 when $b \to \infty$. Since several exact expressions of $Q_\nu(a,b)$ for odd multiple of 0.5 order $\nu$ were obtained in [32]-[33] and [21], $\sqrt{Q_{\nu-0.5}(a,b)Q_{\nu+0.5}(a,b)}$ can be expressed in closed-form. Moreover, our further investigation showed that $\sqrt{Q_{\nu-0.5}(a,b)Q_{\nu+0.5}(a,b)}$ is also a lower bound of $Q_\nu(a,b)$ of integer order $\nu$, which is equivalent to

$$\log Q_\nu(a,b) \geq \frac{1}{2}[\log Q_{\nu-0.5}(a,b) + \log Q_{\nu+0.5}(a,b)],$$
$$\text{for integer } \nu. \quad (5)$$

This motivated us to start working on the log-concavity of the generalized Marcum $Q-$function and the Nuttall $Q-$function.

In [44], we proved the strict log-concavity of the functions $b \mapsto Q_\nu(a,b)$ and $b \mapsto Q_\nu(a,\sqrt{b})$ for all $\nu > 1$ and $a \geq 0$. Moreover, based on some preliminary results, we conjectured that the function $\nu \mapsto Q_\nu(a,b)$ is strictly log-concave on $(0,\infty)$ for all $a \geq 0$ and $b > 0$. In this paper, we are able to verify the above conjecture on $[1,\infty)$. We present a comprehensive study on the log-concavity of the generalized Marcum and Nuttall $Q-$functions. New tight bounds for real order generalized Marcum and Nuttall $Q-$functions are proposed based on the log-concavity of these functions. Our proposed bounds are much tighter than the existing bounds in the literature in most of the cases. They involve only exponential function and the erfc function, and therefore can be computed very efficiently. The relative errors of the proposed bounds converge to 0 when $b \to \infty$. The numerical results show that the absolute relative errors of the proposed bounds are less than 5% in most of the cases. To the extent of the authors' knowledge, our bounds for the generalized Marcum and Nuttall $Q-$functions are the first bounds with such tightness in terms of relative errors on the whole region of $b \in (0,\infty)$.

The detailed content is as follows:

In Section II, we first review the probabilistic interpretations of the generalized Marcum and Nuttall $Q-$functions related to the non-central chi and chi-square distribution, which form the basis of this paper.

Then, in Section III, we present a simple probabilistic method which proves in an unified way the monotonicity of the generalized Marcum and Nuttall $Q-$functions with respect to different parameters.

In Section IV, we first recall a mathematical concept named total positivity, which plays an important role in the following proofs. In complement to the results of Theorem 2.7 in [44], we prove that the functions $b \mapsto 1 - Q_\nu(a,\sqrt{b})$, $b \mapsto 1 - Q_\nu(a,b)$ are log-concave on $(0,\infty)$ for all $\nu \geq 1$ and $\nu \geq 3/2$, respectively. Then, we show that the functions $\nu \mapsto Q_\nu(a,b)$ and $\nu \mapsto 1 - Q_\nu(a,b)$ are log-concave on $[1,\infty)$, while the functions $\nu \mapsto Q_\nu(0,b)$, $\nu \mapsto Q_\nu(a,b) - Q_\nu(0,b)$ and $\nu \mapsto 1 - Q_\nu(0,b)$ are log-concave on $(0,\infty)$, for all admissible values of $a$ and $b$. We also prove that the functions $a \mapsto Q_\nu(\sqrt{a},b)$, $a \mapsto 1 - Q_\nu(\sqrt{a},b)$ and $a \mapsto 1 - Q_\nu(a,b)$ are log-concave on $(0,\infty)$. Some remarks and conjectures are also provided.

In view of the close relationship between the Nuttall

$Q-$function and the generalized Marcum $Q-$function, the log-concavity of the Nuttall $Q-$function and its deformations are established in Section V.

Closed-form lower and upper bounds are proposed for the generalized Marcum and Nuttall $Q-$functions in Section VI. Numerical results and rigorous analysis are also provided to justify the tightness of the bounds.

Some applications of the theoretical results are provided in Section VII.

Finally, the conclusion of this paper is given in Section VIII.

For ease of latter use, we define the following notations:

The probability density function and characteristic function of a random variable $X$ are denoted by $x \mapsto f_X(x)$ and $t \mapsto \varphi_X(t)$, respectively. We use $E_X(X)$ to denote the expectation of $X$, and $E_Y(Y|X=x)$ to denote the conditional expectation of $Y$ given at the value $X = x$. The indicator function of a set $S$ is defined as

$$\mathcal{L}_S(x) = \begin{cases} 1, & \text{if } x \in S, \\ 0, & \text{otherwise.} \end{cases} \quad (6)$$

We use $\mathbb{R}$ to denote the set of real numbers, $\mathbb{N}$ to denote the set of positive integers.

When we discuss the log-concavity of a function $f$, it is convenient to allow $f$ to take on the value zero, in which case we take $\log f(x) = -\infty$ [47], [48]. By this means, we have that the indicator function $\mathcal{L}_S$ of a convex set $S$ is log-concave.

## II. PROBABILISTIC INTERPRETATIONS OF THE GENERALIZED MARCUM AND NUTTALL $Q-$FUNCTIONS

The generalized Marcum $Q-$function $Q_\nu(a,b)$ has an important interpretation in probability theory: it is the complementary cumulative distribution function (ccdf) or reliability function of the normalized non-central chi-square distribution with $2\nu$ degrees of freedom [4, p. 82]. For this, let $X_1, X_2, \ldots, X_\nu$ be random variables that are normally distributed with unit variance and nonzero mean $\gamma_i$, where $i \in \{1, 2, \ldots, \nu\}$. It is known that the random variable $X_1^2 + X_2^2 + \cdots + X_\nu^2$ has the non-central chi-square distribution with $\nu$ degrees of freedom and non-centrality parameter $\lambda = \gamma_1^2 + \gamma_2^2 + \cdots + \gamma_\nu^2$. The probability density function (pdf) $f_{\chi^2_{\nu,\lambda}}: (0, \infty) \to (0, \infty)$ of the non-central chi-square distribution is defined as

$$f_{\chi^2_{\nu,\lambda}}(x) = 2^{-\nu/2} e^{-(x+\lambda)/2} \sum_{k \geq 0} \frac{x^{\nu/2+k-1}(\lambda/4)^k}{\Gamma(\nu/2+k)k!}$$

$$= \frac{e^{-(x+\lambda)/2}}{2}\left(\frac{x}{\lambda}\right)^{\nu/4-1/2} I_{\nu/2-1}\left(\sqrt{\lambda x}\right), \quad (7)$$

where $\Gamma$ is the Euler gamma function [49, eq. (6.1.1)]. Although $\nu$ is an integer in our description of the non-central chi-square distribution above, it is known that the distribution, defined by (7), is a proper distribution for any positive $\nu$ [46, p. 436] and [44]. Consequently, the generalized Marcum $Q-$function can be expressed as

$$Q_\nu\left(\sqrt{a}, \sqrt{b}\right) = \int_b^\infty f_{\chi^2_{2\nu,a}}(x)dx$$

$$= \int_b^\infty \frac{1}{2}\left(\frac{x}{a}\right)^{\nu/2-1/2} e^{-(x+a)/2} I_{\nu-1}\left(\sqrt{ax}\right) dx. \quad (8)$$

In view of the similarity of the definitions of the generalized Marcum and Nuttall $Q-$functions (1) and (2), one can find that the normalized Nuttall $Q-$function is actually the upper-side partial (truncated) moment of the normalized non-central chi-square distribution, which is given by

$$\mathcal{Q}_{\mu,\nu}\left(\sqrt{a}, \sqrt{b}\right) = \int_b^\infty x^{(\mu-\nu-1)/2} f_{\chi^2_{2(\nu+1),a}}(x)dx$$

$$= \int_b^\infty \frac{x^{(\mu-1)/2}}{2a^{\nu/2}} e^{-(x+a)/2} I_\nu\left(\sqrt{ax}\right) dx. \quad (9)$$

If $\lambda = 0$, the above distribution reduces to the classical (central) chi-square distribution, whose pdf is

$$f_{\chi^2_\nu}(x) = f_{\chi^2_{\nu,0}}(x) = \frac{x^{\nu/2-1}e^{-x/2}}{2^{\nu/2}\Gamma(\nu/2)}. \quad (10)$$

Consequently, (8) reduces to

$$Q_\nu\left(0, \sqrt{b}\right) = \int_b^\infty f_{\chi^2_{2\nu}}(x)dx = \int_b^\infty \frac{x^{\nu-1}e^{-x/2}}{2^\nu \Gamma(\nu)} dx. \quad (11)$$

We note that if the random variables $\chi^2_{\nu_1,\lambda_1}$ and $\chi^2_{\nu_2,\lambda_2}$, which have non-central chi-square distribution with $\nu_1$, $\nu_2$ degrees of freedom, respectively, and non-centrality parameters $\lambda_1$, $\lambda_2$, respectively, are independent, then the random variable $\chi^2_{\nu_1,\lambda_1} + \chi^2_{\nu_2,\lambda_2}$ has a non-central chi-square distribution with $\nu_1 + \nu_2$ degrees of freedom and non-centrality parameter $\lambda_1 + \lambda_2$. Therefore, the non-central $\chi^2$ distributions is reproductive under convolution [46]. This can be explained by the characteristic functions of the non-central and central chi-square distributions, which are given by

$$\varphi_{\chi^2_{\nu,\lambda}}(t) = e^{\frac{i\lambda t}{1-2it}}(1-2it)^{-\nu/2}, \quad (12)$$

and

$$\varphi_{\chi^2_\nu}(t) = (1-2it)^{-\nu/2}. \quad (13)$$

It may be a little surprising that the degrees of freedom $\nu$ can be 0 in (12), and the classical non-central $\chi^2$ distribution reduces to non-central $\chi^2$ distributions of zero degrees of freedom, i.e. $\chi^2_{0,\lambda}$ [50]. $\chi^2_{0,\lambda}$ can be also defined as a mixture of the distributions $0, \chi^2_2, \chi^2_4, \ldots$ with Poisson weights. The above mentioned reproductive property holds true for non-central $\chi^2$ distributions of zero degrees of freedom, i.e. $\chi^2_{0,\lambda_1} + \chi^2_{0,\lambda_2} \sim \chi^2_{0,\lambda_1+\lambda_2}$, $\chi^2_{0,\lambda} + \chi^2_\nu \sim \chi^2_{\nu,\lambda}$ and $\chi^2_{0,\lambda_1} + \chi^2_{\nu,\lambda_2} \sim \chi^2_{\nu,\lambda_1+\lambda_2}$. However, the random variable $\chi^2_{0,\lambda}$ does not possess a probability density function, because of a discrete mass of $e^{-\frac{1}{2}\lambda}$ at the zero point, and the positive part of this distribution has a density function $f_{\chi^2_{0,\lambda}}$, by excluding the probability at $x = 0$, which can be expressed as

$$f_{\chi^2_{0,\lambda}}(x) = \frac{1}{x}e^{-(\lambda+x)/2} \sum_{k=1}^\infty \frac{(\frac{1}{4}\lambda x)^k}{k!(k-1)!}$$

$$= \frac{1}{2}\left(\frac{\lambda}{x}\right)^{1/2} e^{-(\lambda+x)/2} I_1\left(\sqrt{\lambda x}\right).^1 \quad (14)$$

The ccdf of $\chi^2_{0,\lambda}$ is

$$P\left(\chi^2_{0,\lambda} \geq b\right)
= \begin{cases} \int_b^\infty \frac{1}{2}\left(\frac{\lambda}{x}\right)^{1/2} e^{-\frac{x+\lambda}{2}} I_1\left(\sqrt{\lambda x}\right) dx, & \text{if } b > 0, \\ 1, & \text{if } b \leq 0, \end{cases} \quad (15)$$

which has a discontinuity point at $b = 0$.

There are also alternative probabilistic interpretations of the generalized Marcum and Nuttall $Q-$functions related to the non-central chi distribution. It is known that if $Y_1, Y_2, \ldots, Y_\nu$ are random variables that are normally distributed with unit variance and nonzero mean $\mu_1, \mu_2, \ldots, \mu_\nu$, then the random variable $[Y_1^2 + Y_2^2 + \ldots + Y_\nu^2]^{1/2}$ has the non-central chi distribution with $\nu$ degrees of freedom and non-centrality parameter $\tau = [\mu_1^2 + \mu_2^2 + \ldots + \mu_\nu^2]^{1/2}$. The pdf $f_{\chi_{\nu,\tau}} : (0, \infty) \to (0, \infty)$ of the non-central chi distribution [46] is defined as

$$f_{\chi_{\nu,\tau}}(x) = 2^{-\frac{\nu}{2}+1} e^{-\frac{x^2+\tau^2}{2}} \sum_{k \geq 0} \frac{x^{\nu+2k-1}(\tau/2)^{2k}}{\Gamma(\nu/2+k)k!}$$
$$= \tau e^{-\frac{x^2+\tau^2}{2}} \left(\frac{x}{\tau}\right)^{\nu/2} I_{\frac{\nu}{2}-1}(\tau x), \quad (16)$$

where $\nu$ can be also an arbitrary positive real number. Therefore, the generalized Marcum $Q-$function and normalized Nuttall $Q-$function can be expressed as

$$Q_\nu(a, b) = \int_b^\infty f_{\chi_{2\nu,a}}(x) dx, \quad (17)$$

and

$$\mathcal{Q}_{\mu,\nu}(a, b) = \int_b^\infty x^{\mu-\nu-1} f_{\chi_{2(\nu+1),a}}(x) dx, \quad (18)$$

respectively. If $\tau = 0$, the above distribution reduces to the classical chi distribution with pdf $f_{\chi_\nu} : (0, \infty) \to (0, \infty)$, defined by

$$f_{\chi_\nu}(x) = f_{\chi_{\nu,0}}(x) = \frac{x^{\nu-1} e^{-x^2/2}}{2^{\nu/2-1}\Gamma(\nu/2)}. \quad (19)$$

Therefore, (17) reduces to

$$Q_\nu(0, b) = \int_b^\infty f_{\chi_{2\nu}}(x) dx = \int_b^\infty \frac{x^{2\nu-1} e^{-x^2/2}}{2^{\nu-1}\Gamma(\nu)} dx. \quad (20)$$

Some more probabilistic interpretations of $Q_\nu(a, b)$ can be found in [10], [32] and [51]-[52].

The chi and chi-square distributions have close connection to a number of useful distributions in digital communications, such as Rayleigh distribution, Rician distribution, Nakagami-$m$ distribution, generalized Rayleigh and Rician distribution, bivariate Rayleigh and Nakagami-$m$ distribution, exponential distribution, gamma distribution, etc. [3]-[4], [53]-[54]. This is exactly the reason for the vast applications of the generalized Marcum and Nuttall $Q-$functions in wireless communications.

[1]There is a mistake in the original formula given in Siegel's paper [50].

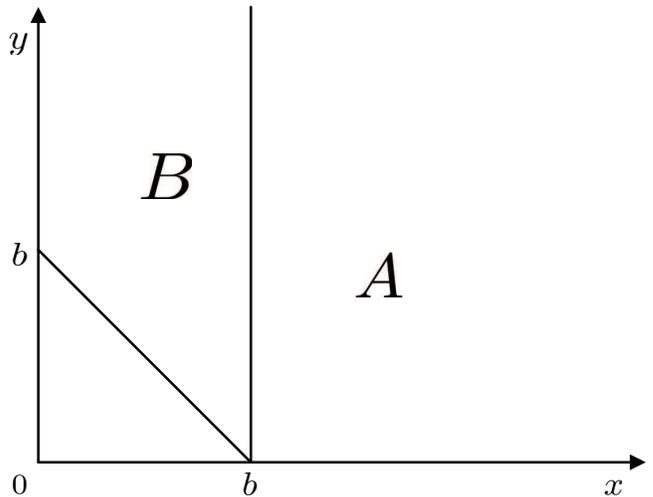

Fig. 1. A geometrical intuition of our proof idea of Lemma 1.

### III. THE MONOTONICITY OF THE GENERALIZED MARCUM AND NUTTALL $Q-$FUNCTIONS

The monotonicity of the generalized Marcum and Nuttall $Q-$functions has been discussed via a number of analytical proofs [21], [32], [43], [46]. However, these proofs did not provide much insight about the functions, and each of proof could only show the monotonicity with respect to one parameter.

In this section, we utilize the powerful probabilistic interpretations of the generalized Marcum and Nuttall $Q-$functions that we have discussed earlier, and present a simple probabilistic method to solve in an unified way the monotonicity of these functions with respect to different parameters. A geometrical intuition of our method is also provided to facilitate the understanding of the proof.

In part (a) of Theorem 3.1 in [44], we have used a simple probabilistic result for independent random variables $X$ and $Y$ with non-negative support, i.e.

$$P(X + Y \geq b) > P(X \geq b), \text{ for } b > 0, \quad (21)$$

to prove that $\nu \mapsto Q_\nu(a, b)$ is strictly increasing for $a \geq 0$ and $b > 0$.

The inequality (21) can be explained simply in a geometrical way, as illustrated in Fig. 1. We consider the first quadrant in the coordinate plane of $x$ and $y$, where $A$ is the region $\{(x,y)|x \geq b, y \geq 0\}$ and $B$ is the region $\{(x,y)|x+y \geq b, 0 \leq x < b, y \geq 0\}$. Then, $P(X + Y \geq b)$ is the sum of the probabilistic integrations on the regions $A$ and $B$, while $P(X \geq b)$ is the probabilistic integration only on the region $A$. By this, (21) is expected.

In order to solve in an unified way the monotonicity of the generalized Marcum and Nuttall $Q-$functions, we now generalize the inequality (21) by inducing an non-decreasing positive weight function in the probabilistic integrations. Our new result is stated as

**Lemma 1.** *Let $X$ and $Y$ be independent random variables with non-negative support and let $g : (0, \infty) \to (0, \infty)$*

be an non-decreasing positive function. Further, let $f_X$ and $f_{X+Y}$ be the pdfs of the random variables $X$ and $X + Y$, respectively. Let $F_Y$ be the cumulative distribution function (cdf) of the random variable $Y$, which may not has a pdf if $F_Y$ is discontinuous. Then, the following inequality is true for each $b > 0$

$$\int_b^\infty g(t) f_{X+Y}(t) dt > \int_b^\infty g(t) f_X(t) dt, \quad (22)$$

if $F_Y(0) < 1$ and the integrals exist.

The proof of Lemma 1 is given in Appendix A, which also fits the geometrical intuition of Fig. 1.

With this Lemma, we now can prove the strict monotonicity of the generalized Marcum and Nuttall $Q-$functions.

We first consider the case when $g(x) = 1$ for all $x \in (0, \infty)$. If $X \sim \chi^2_{2\nu_1, a}$ and $Y \sim \chi^2_{2\nu_2}$, then clearly $X + Y \sim \chi^2_{2(\nu_1+\nu_2), a}$. Hence, by Lemma 1 we easily have that (see also Theorem 3.1 in [44]) $\nu \mapsto Q_\nu(\sqrt{a}, \sqrt{b})$ is strictly increasing on $(0, \infty)$ for each $a \geq 0$ and $b > 0$, i.e. we have

$$Q_{\nu_1+\nu_2}\left(\sqrt{a}, \sqrt{b}\right) > Q_{\nu_1}\left(\sqrt{a}, \sqrt{b}\right) \quad (23)$$

for all $\nu_1, \nu_2, b > 0$ and $a \geq 0$. On the other hand if $X \sim \chi^2_{2\nu, a_1}$ and $Y \sim \chi^2_{0, a_2}$, then it is easy to see that $X + Y \sim \chi^2_{2\nu, a_1+a_2}$. We note that $\chi^2_{0, a_2}$ possesses a discontinuous cdf, and therefore has no pdf. Consequently, in view of Lemma 1 we obtain that the function $a \mapsto Q_\nu(\sqrt{a}, \sqrt{b})$ is strictly increasing on $[0, \infty)$ for each $\nu, b > 0$, i.e.

$$Q_\nu\left(\sqrt{a_1 + a_2}, \sqrt{b}\right) > Q_\nu\left(\sqrt{a_1}, \sqrt{b}\right) \quad (24)$$

for all $a_1 \geq 0$ and $a_2, \nu, b > 0$.

Finally, let $g(x) = x^{\mu/2}$ for all $x \in (0, \infty)$ and $\mu \geq 0$. A similar argument as we presented above yield the following inequalities for the normalized Nuttall $Q-$function

$$\mathcal{Q}_{\mu+\nu_1+\nu_2, \nu_1+\nu_2-1}\left(\sqrt{a}, \sqrt{b}\right) > \mathcal{Q}_{\mu+\nu_1, \nu_1-1}\left(\sqrt{a}, \sqrt{b}\right) \quad (25)$$

and

$$\mathcal{Q}_{\mu+\nu, \nu-1}\left(\sqrt{a_1+a_2}, \sqrt{b}\right) > \mathcal{Q}_{\mu+\nu, \nu-1}\left(\sqrt{a_1}, \sqrt{b}\right) \quad (26)$$

where $\nu_1, \nu \geq 1$, $\mu \geq 0$ and $a_1, a_2, \nu_2, a, b > 0$. Moreover, it is easy to verify that the generalized Marcum and Nuttall $Q-$functions are strictly decreasing with respect to $b$ on $[0, \infty)$. These results are summarized in the following theorem.

**Theorem 1.** *The following assertions are true:*

(a) *The generalized Marcum $Q-$function $Q_\nu(a, b)$ is strictly increasing in $\nu$ and $a$ for all $a \geq 0$ and $b, \nu > 0$, and is strictly decreasing in $b$ for all $a, b \geq 0$ and $\nu > 0$.*
(b) *The normalized Nuttall $Q-$function $\mathcal{Q}_{\mu+\nu+1, \nu}(a, b)$ is strictly increasing in $\nu$ and $a$ for all $\nu, \mu \geq 0$ and $a, b > 0$, and is strictly decreasing in $b$ for all $b, \nu, \mu \geq 0$ and $a > 0$.*
(c) *The standard Nuttall $Q-$function $Q_{\mu+\nu+1, \nu}(a, b)$ is strictly increasing in $\nu$ for all $a \geq 1$, $\nu, \mu \geq 0$ and $b > 0$, is strictly increasing in $a$ for all $\nu, \mu \geq 0$ and $a, b > 0$, and is strictly decreasing in $b$ for all $b, \nu, \mu \geq 0$ and $a > 0$.*

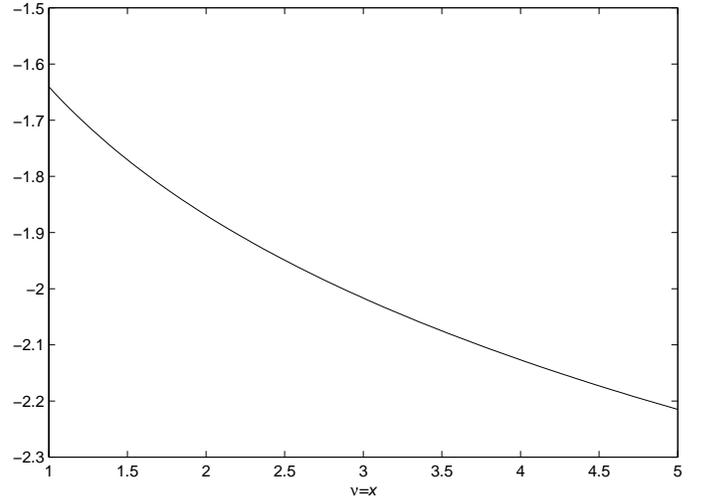

Fig. 2. The logarithm of the pdf of non-central chi-square distribution, i.e. $\log f_{\chi^2_{\nu,\lambda}}(x)$, when $\lambda = 2$ and $\nu = x$.

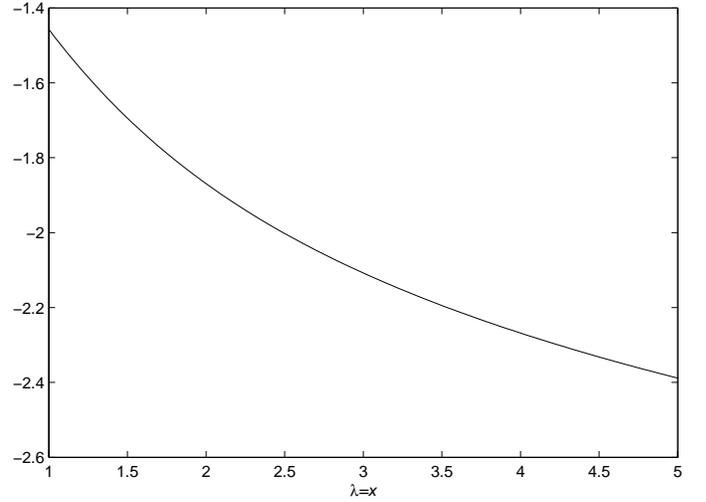

Fig. 3. The logarithm of the pdf of non-central chi-square distribution, i.e. $\log f_{\chi^2_{\nu,\lambda}}(x)$, when $\nu = 2$ and $\lambda = x$.

Therefore, we have obtained the monotonicity of the generalized Marcum and Nuttall $Q-$functions with respect to variables $a, b$ and $\nu$ with an unified and rather simple probabilistic method. To the best of the authors' knowledge, the monotonicity of the standard and normalized Nuttall $Q-$functions with respect to the non-centrality parameter $a$ has not been discussed in the literature.

## IV. THE LOG-CONCAVITY OF THE GENERALIZED MARCUM $Q-$FUNCTION

In [44], Sun and Baricz conjectured that the function $\nu \mapsto Q_\nu(a, b)$ is strictly log-concave on $(0, \infty)$ for all $a \geq 0$ and $b > 0$. In order to prove this, we have tried several methods. One idea is to consider the generalized Marcum $Q-$function as a function of two variables and to prove that it is log-concave as a function of these two variables. It is known that the pdf of the non-central chi-square distribution $f_{\chi^2_{2\nu, a}}$ is log-concave with respect to $\nu$ and is also log-concave with respect to $a$ under some requirements [55]. However, these results do not help too much in the study of the log-concavity of the functions $\nu \mapsto Q_\nu(a,b)$ and $a \mapsto Q_\nu(\sqrt{a},b)$, because we may need the function $(\nu,x) \mapsto f_{\chi^2_{\nu,\lambda}}(x)$ or $(\lambda,x) \mapsto f_{\chi^2_{\nu,\lambda}}(x)$ to be log-concave on two variables [47, p. 106][1], which is not true. In Fig. 2 and 3, we show the numerical results of the functions $x \mapsto \log f_{\chi^2_{x,2}}(x)$ and $x \mapsto \log f_{\chi^2_{2,x}}(x)$, from which one can understand that $(\nu,x) \mapsto f_{\chi^2_{\nu,\lambda}}(x)$ or $(\lambda,x) \mapsto f_{\chi^2_{\nu,\lambda}}(x)$ can not be log-concave.

Fortunately, we have found a powerful mathematical concept, named total positivity, which can help us to conquer this difficulty. For the reader's convenience, we first offer a brief introduction to total positivity, which is necessary for our proofs. More details about total positivity can be found in Karlin's monograph [57].

### A. A brief introduction to total positivity

The definition of totally positive function [57, p. 11] is the following:

**Definition 1.** *A function $f(x,y)$ of two real variables, $x$ ranging over $\Delta_1$ and $y$ ranging over $\Delta_2$, is said to be totally positive of order $r$ ($TP_r$), if for all $x_i \in \Delta_1$ and $y_i \in \Delta_2$, $i \in \{1,2,\ldots,m\}$ such that $x_1 < x_2 < \cdots < x_m$, $y_1 < y_2 < \cdots < y_m$, where $m \in \{1,2,\ldots,r\}$, we have the inequalities*

$$f\begin{pmatrix} x_1, x_2, \ldots, x_m \\ y_1, y_2, \ldots, y_m \end{pmatrix}$$
$$= \begin{vmatrix} f(x_1,y_1) & f(x_1,y_2) & \ldots & f(x_1,y_m) \\ \vdots & \vdots & \vdots & \vdots \\ f(x_m,y_1) & f(x_m,y_2) & \ldots & f(x_m,y_m) \end{vmatrix} \geq 0. \quad (27)$$

Either of $\Delta_1$ and $\Delta_2$ can be an interval of real line, or a sequence of countable discrete points.

A related concept to total positivity is the sign reverse regularity.

**Definition 2.** *A function $f(x,y)$ is said to be sign reverse regular of order $k$ ($SRR_k$), if for every $x_1 < x_2 < \cdots < x_m$ and $y_1 < y_2 < \cdots < y_m$, where $m \in \{1,2,\ldots,k\}$, the sign of*

$$f\begin{pmatrix} x_1, x_2, \ldots, x_m \\ y_1, y_2, \ldots, y_m \end{pmatrix} \quad (28)$$

*is $(-1)^{m(m-1)/2}$.*

If a function $f(x,y)$ is said to be $TP_\infty$, if it is $TP_k$ for all $k \in \mathbb{N}$. Similarly, it is said to be $SRR_\infty$, if it is $SRR_k$ for all $k \in \mathbb{N}$.

Totally positive functions can be generated by the following composition law, which expresses a continuous analogue for evaluating determinants arising in matrix multiplication [57, pp. 16-17].

---
[1] If $f(x,z)$ is log-convex in $z$ for each $x \in C$, then $g(z) = \int_C f(x,z)dx$ is log-convex, where $C$ is an arbitrary set of $x$ [47, p. 106] and [56, p. 9]. However, a similar statement of this does not hold true for log-concavity. Instead, the following statement is true for log-concavity: If $f(x,z) : \mathbb{R}^n \times \mathbb{R}^m \mapsto \mathbb{R}$ is log-concave in $x$ and $z$, then $g(z) = \int f(x,z)dx$ is a log-concave function of $z$ on $\mathbb{R}^m$ [47, p. 106] and [48, Theorem 3].

**Lemma 2.** *If $r(x,w) = \int p(x,t)q(t,w)d\sigma(t)$ and the integral converges absolutely, then*

$$r\begin{pmatrix} x_1, x_2, \ldots, x_k \\ w_1, w_2, \ldots, w_k \end{pmatrix} = \int\int\cdots\int_{t_1<t_2<\cdots<t_k}$$
$$p\begin{pmatrix} x_1, x_2, \ldots, x_k \\ t_1, t_2, \ldots, t_k \end{pmatrix} q\begin{pmatrix} t_1, t_2, \ldots, t_k \\ w_1, w_2, \ldots, w_k \end{pmatrix}$$
$$d\sigma(t_1)d\sigma(t_2)\ldots d\sigma(t_k). \quad (29)$$

Here, $d\sigma(t)$ denotes a sigma-finite measure defined on $t \in \Delta$. When $\Delta$ consists of a discrete set, the integral is interpreted as a sum.

The above result has an immediate corollary, expressed as follows [57]

**Lemma 3.** *If $p(x,t)$ is $TP_m$ and $q(t,w)$ is $TP_n$, then $r(x,w) = \int p(x,t)q(t,w)d\sigma(t)$ is $TP_{\min\{m,n\}}$, provided that $\sigma(t)$ is a regular finite measure; when $p(x,t)$ is $SRR_m$ and $q(t,w)$ is $TP_n$, then $r(x,w)$ is $SRR_{\min\{m,n\}}$.*

Analogously, as in Lemma 2, when $t$ is chosen from a discrete set, the above integral is interpreted as a sum.

The following lemma states another important composition law of total positive functions [57, p. 130].

**Lemma 4.** *Let $f(x+y)$ be $SRR_r$ for $x,y > 0$. Suppose that $\phi(t,x)$ is $TP_r$ for $t,x > 0$, and satisfies*

$$\phi(t+s,x) = \int_0^x \phi(t,\xi)\phi(s,x-\xi)d\xi. \quad (30)$$

*If $c$ is defined by*

$$c(t) = \int_0^\infty \phi(t,x)f(x)dx, \quad (31)$$

*then $c(t+s)$ is $SRR_r$ for $t,s > 0$.*

We note that formula (30) expresses exactly the reproductive property of $\phi(t,x)$ under convolution.

Some other results which are useful in the sequel are included in the following lemma. The proof of parts (a)-(d) can be found in [57] and part (e) is stated in [47, p. 79].

**Lemma 5.** *The following assertions are true:*

(a) $f(x,y) = x^y$ is $TP_2$ for $\Delta_1 = (0,\infty)$, $\Delta_2 = (-\infty,\infty)$.
(b) *If $f(x,y)$ is $TP_r$, let $\phi(x), \varphi(y)$ maintain the same constant sign on $\Delta_1$ and $\Delta_2$, respectively, then $\phi(x)\varphi(y)f(x,y)$ is also $TP_r$.*
(c) *Let $f$ be a strictly positive second order differentiable function. Then $(x,y) \mapsto f(x-y)$ is $TP_2$, if and only if $x \mapsto f(x)$ is log-concave on its domain.*
(d) $(x,y) \mapsto f(x+y)$ is $SRR_2$, if and only if $x \mapsto f(x)$ is also log-concave on its domain.
(e) *Let $f : \mathbb{R}^n \mapsto \mathbb{R}$, $A \in \mathbb{R}^{n\times m}$, and $b \in \mathbb{R}^n$, and consider the function $g : \mathbb{R}^m \mapsto \mathbb{R}$, defined by*

$$g(x) = f(Ax+b), \quad (32)$$

*with $\mathrm{dom} g = \{x \mid Ax+b \in \mathrm{dom} f\}$. If $f$ is log-concave, then $g$ is log-concave too.*

## B. The log-concavity of $Q_\nu(a,b)$ with respect to $b$

Recently, we proved in [44] that the functions $b \mapsto Q_\nu(a,\sqrt{b})$ and $b \mapsto Q_\nu(a,b)$ are both strictly log-concave on $(0,\infty)$ for all $a \geq 0$ and $\nu > 1$. Now we give the complementary results as follows.

**Theorem 2.** *If $a \geq 0$, then the functions $b \mapsto 1 - Q_\nu(a,\sqrt{b})$ and $b \mapsto 1 - Q_\nu(a,b)$ are log-concave on $(0,\infty)$ for all $\nu \geq 1$ and $\nu \geq 3/2$, respectively.*

*Proof:* The function $1 - Q_\nu(a,\sqrt{b})$ is the cdf of non-central chi-square distribution, given by

$$1 - Q_\nu(a,\sqrt{b}) = \int_0^b f_{\chi^2_{2\nu,a}}(x)dx. \qquad (33)$$

We know that the function $x \mapsto f_{\chi^2_{2\nu,a}}(x)$ is log-concave on $(0,\infty)$ for all $a \geq 0$ and $\nu \geq 1$ [55]. On the other hand it is known that if the probability density function is log-concave, the cumulative distribution function is also log-concave (see [58] and the references therein). Therefore, $b \mapsto 1 - Q_\nu(a,\sqrt{b})$ is log-concave too on $(0,\infty)$ for all $a \geq 0$ and $\nu \geq 1$.

We have proved in [44] that the pdf of non-central chi distribution $b \mapsto f_{\chi_{2\nu,a}}(b)$ is log-concave on $(0,\infty)$ for all $a > 0$ and $\nu \geq 3/2$. When $a = 0$, the pdf of central chi distribution $b \mapsto f_{\chi_{2\nu}}(b)$, defined by (19), is log-concave on $(0,\infty)$ for all $\nu \geq 1/2$. From (17) and (20), we have

$$1 - Q_\nu(a,b) = \int_0^b f_{\chi_{2\nu,a}}(x)dx. \qquad (34)$$

By using the same method as above, we conclude that the function $b \mapsto 1 - Q_\nu(a,b)$ is log-concave on $(0,\infty)$ for all $a \geq 0$ and $\nu \geq 3/2$, and the proof is complete. ∎

We have shown the log-concavity of the Marcum $Q-$function and its deformations with respect to $b$ for not quite small $\nu$. Now we discuss the cases of small $\nu$ which is suggested by our numerical results.

*Remark 1:* It was shown that $b \mapsto Q_\nu(0,\sqrt{b})$ is log-concave for $\nu \in [1,\infty)$ and log-convex for $\nu \in (0,1]$ in [59]. On the other hand, our numerical results show that $b \mapsto Q_\nu(a,\sqrt{b})$ is neither log-convex nor log-concave if $\nu \in (0,1]$ and $a > 0$. Therefore, our previous result that $b \mapsto Q_\nu(a,\sqrt{b})$ is strictly log-concave on $(0,\infty)$ for all $a \geq 0$ and $\nu > 1$ is sharp.

On the other hand, it was shown that the pdf of chi distribution is log-concave for $\nu \geq 1$, which indicates that $b \mapsto Q_\nu(0,b)$ is log-concave on $(0,\infty)$ for all $\nu \geq 1/2$ [59]. our numerical results suggest that $b \mapsto Q_\nu(a,b)$ is log-concave on $(0,\infty)$ for all $a > 0$ and $\nu \geq 1/2$. Therefore, we conjecture that

**Conjecture 1.** *The function $b \mapsto Q_\nu(a,b)$ is log-concave on $(0,\infty)$ for all $a > 0$ and $\nu \in [1/2, 1]$.*

A special case of Conjecture 1 is when $\nu = 1/2$. In this case, the conjecture reduces to $b \mapsto Q_\nu(a,b) = Q(b+a) + Q(b-a)$ is log-concave on $(0,\infty)$ for $a \geq 0$.

*Remark 2:* In [59], Bagnoli and Bergstrom proved that the cdf of gamma distribution is a log-concave function. Since the chi-square distribution with $\nu$ degrees of freedom is a gamma distribution with parameter $\nu/2$, they derived that $b \mapsto 1 - Q_\nu(0,\sqrt{b})$ is log-concave for all $\nu \geq 0$ [59, p. 16]. Moreover, using the same proof idea for the case of chi distribution, one can show that $b \mapsto 1 - Q_\nu(0,b)$ is log-concave for all $\nu \geq 0$. We now provide a conjecture for the case $a \geq 0$, stated as

**Conjecture 2.** *If $a > 0$, then the functions $b \mapsto 1 - Q_\nu(a,\sqrt{b})$ and $b \mapsto 1 - Q_\nu(a,b)$ are log-concave on $(0,\infty)$ for all $\nu \in [0,1)$ and $\nu \in [1/2, 3/2)$, respectively.*

We note that our numerical results suggest that $b \mapsto 1 - Q_0(a,\sqrt{b})$ is log-concave on $(0,\infty)$ while $b \mapsto 1 - Q_\nu(a,b)$ is not log-concave for $\nu \in (0,1/2)$.

## C. The log-concavity of $Q_\nu(a,b)$ with respect to $\nu$

In [44], we deduced a Turán type inequality, which is interesting in its own right [55], given as

$$Q_{\nu+1}^2(a,b) \geq Q_\nu(a,b)Q_{\nu+2}(a,b). \qquad (35)$$

It is known that the integrand of $Q_\nu(\sqrt{a},\sqrt{b})$ as a function of $\nu$, i.e. $\nu \mapsto f_{\chi^2_{2\nu,a}}(x)$ is log-concave on $(0,\infty)$ [55]. Thus, the above Turán type inequality suggests that this log-concavity property remains true after integration, of course with some assumptions on parameters. Taking into account this observation in [44] we conjectured that the function $\nu \mapsto Q_\nu(a,b)$ is actually strictly log-concave on $(0,\infty)$ for all $a \geq 0$ and $b > 0$. Now, we are able to verify this conjecture for $\nu \geq 1$. The case $\nu \in (0,1]$ remains open.

**Theorem 3.** *The following assertions are true:*
(a) *The function $\nu \mapsto Q_\nu(0,b)$ is log-concave on $(0,\infty)$ for all $b \geq 0$.*
(b) *The function $\nu \mapsto Q_\nu(a,b)$ is log-concave on $[1,\infty)$ for all $a,b \geq 0$.*
(c) *The function $\nu \mapsto Q_\nu(a,b) - Q_\nu(0,b)$ is log-concave on $(0,\infty)$ for all $a,b > 0$.*
(d) *The function $\nu \mapsto 1 - Q_\nu(0,b)$ is log-concave on $[0,\infty)$ for all $b > 0$.*
(e) *The function $\nu \mapsto 1 - Q_\nu(a,b)$ is log-concave on $[1,\infty)$ for all $a \geq 0$ and $b > 0$.*

The proof of Theorem 3 is provided in Appendix B.

*Remark 3:* It is worth to mention here that very recently Alzer and Baricz [60] by using an interesting idea of Alzer [61] proved that for all $b > 0$ fixed the function $\nu \mapsto Q_\nu(0,\sqrt{2b})$ is log-concave on $(0,\infty)$. Clearly this implies that for all $b \geq 0$ the function $\nu \mapsto Q_\nu(0,b)$ is log-concave too on $(0,\infty)$, which is exactly the statement of part (a) of Theorem 3. However, in Appendix B we present a completely different proof for this part.

Moreover, we note that Merkle [62] based on a Turán-type inequality involving the incomplete gamma function conjectured that the function $\nu \mapsto 1 - Q_\nu(0,\sqrt{2b})$ is log-concave on $(0,\infty)$ for all $b > 0$. A proof of this conjecture can be found in Alzer's paper [61]. Clearly, part (d) of Theorem 3 verifies also this conjecture, and the proof given in Appendix B is completely different than in [61].

*Remark 4:* Parts (b) and (e) of Theorem 3 do not include the region $\nu \in (0,1]$. Therefore, we have the following conjecture

**Conjecture 3.** *If $a > 0$ and $b \geq 0$, then the functions $\nu \mapsto Q_\nu(a,b)$ and $\nu \mapsto 1 - Q_\nu(a,b)$ are log-concave on $(0,1]$.*

It is convenient to understand why our proof fails in these cases. The key proof idea of Theorem 3 is to use Lemma 4, which requires the concerned function with a lower order $\nu$ to be log-concave in $b$. We mentioned in *Remark 1* that our numerical results show that $b \mapsto Q_\nu(a, \sqrt{b})$ is neither log-concave nor log-convex for $\nu \in (0,1]$. Therefore, our proposed method can not be used to prove that $\nu \mapsto Q_\nu(a,b)$ is log-concave on $(0,1]$. Some other methods are needed to prove our conjecture.

On the other hand, we have mentioned in *Remark 2* that our numerical results suggest $b \mapsto 1 - Q_0\left(a, \sqrt{b}\right)$ to be log-concave on $(0, \infty)$. If one can prove this, our proof method can derive that Part (e) of Theorem 3 is true on $(0, \infty)$.

### D. The log-concavity of $Q_\nu(a,b)$ with respect to $a$

Now, we study the log-concavity of the generalized Marcum $Q-$function with respect to $a$. It is interesting that the log-concavity holds true for the following integral of $f_{\chi^2_{2\nu,a}}(x)$ on an interval $(c,d) \subseteq (0, \infty)$.

**Lemma 6.** *The function $a \mapsto \int_c^d f_{\chi^2_{2\nu,a}}(x) dx$ is log-concave on $[0,\infty)$ for all $\nu > 0$ and $(c,d) \subseteq (0,\infty)$.*

The proof of Lemma 6 is given in Appendix C.

Our main result of this subsection, which is an immediate application of Lemma 6, reads as follows.

**Theorem 4.** *Let $\nu > 0$. Then the following assertions are true:*
(a) *The function $a \mapsto Q_\nu(\sqrt{a}, b)$ is log-concave on $[0,\infty)$ for all $b \geq 0$.*
(b) *The function $a \mapsto 1 - Q_\nu(\sqrt{a}, b)$ is log-concave on $[0,\infty)$ for all $b > 0$.*
(c) *The function $a \mapsto 1 - Q_\nu(a, b)$ is log-concave on $[0,\infty)$ for all $b > 0$.*

*Proof:* (a) & (b) Using Lemma 6 for the case when $c$ tends to $b^2$, and $d$ tends to $\infty$ we obtain the result of part (a). Similarly, substituting $c$ with $0$ and $d$ with $b^2$ in Lemma 6, part (b) is also proved.

(c) It is known that if a positive function $f$ is log-concave and decreasing, and $g$ is convex, then the composite function $f \circ g$ is log-concave too [47, p. 84]. We choose $f(a) = 1 - Q_\nu(\sqrt{a}, b)$ and $g(a) = a^2$. We known that $a \mapsto f(a)$ is log-concave and strictly decreasing, $a \mapsto g(a)$ is convex, thus $a \mapsto (f \circ g)(a) = 1 - Q_\nu(a,b)$ is also log-concave, which completes the proof of part (c). ∎

## V. THE LOG-CONCAVITY OF THE STANDARD AND NORMALIZED NUTTALL $Q-$FUNCTIONS

The standard and normalized Nuttall $Q-$functions have similar probabilistic interpretations with the generalized Marcum $Q-$function. Therefore, we can establish the log-concavity of the Nuttall $Q-$function and its deformations similarly as in the previous section.

### A. The log-concavity of the Nuttall $Q-$function with respect to $b$

The following theorem presents the log-concavity of the Nuttall $Q-$function with respect to $b$, which is similar with the results of [44, Theorem 2.7] and Theorem 2 of the present paper.

**Theorem 5.** *Let $b, \mu, \nu \geq 0$ and $a > 0$. Then the following assertions about the normalized Nuttall Q-function, $\mathcal{Q}_{\mu,\nu}(a,b)$, are true:*
(a) *The function $b \mapsto \mathcal{Q}_{\mu,\nu}(a, \sqrt{b})$ is log-concave on $[0, \infty)$ for all $\mu + \nu \geq 1$.*
(b) *The function $b \mapsto \mathcal{Q}_{\mu,\nu}(a, b)$ is log-concave on $[0, \infty)$ for all $\mu + \nu \geq 1$.*
(c) *The function $b \mapsto E_{\chi_{2(\nu+1),a}}(X^{\mu-\nu-1}) - \mathcal{Q}_{\mu,\nu}(a, \sqrt{b})$ is log-concave on $(0, \infty)$ for all $\mu + \nu \geq 1$.*
(d) *The function $b \mapsto E_{\chi_{2(\nu+1),a}}(X^{\mu-\nu-1}) - \mathcal{Q}_{\mu,\nu}(a, b)$ is log-concave on $(0, \infty)$ for all $\mu \geq 1$, $\nu \geq 1/2$.*

The proof of Theorem 5 is given in Appendix D. These results can be rewritten in terms of the standard Nuttall $Q-$function as follows.

**Corollary 1.** *Let $b, \mu, \nu \geq 0$ and $a > 0$. Then the following assertions about the standard Nuttall Q-function, $Q_{\mu,\nu}(a,b)$, are true:*
(a) *The function $b \mapsto Q_{\mu,\nu}(a, \sqrt{b})$ is log-concave on $[0, \infty)$ for all $\mu + \nu \geq 1$.*
(b) *The function $b \mapsto Q_{\mu,\nu}(a, b)$ is log-concave on $[0, \infty)$ for all $\mu + \nu \geq 1$.*
(c) *The function $b \mapsto a^\nu E_{\chi_{2(\nu+1),a}}(X^{\mu-\nu-1}) - Q_{\mu,\nu}(a, \sqrt{b})$ is log-concave on $(0, \infty)$ for all $\mu + \nu \geq 1$.*
(d) *The function $b \mapsto a^\nu E_{\chi_{2(\nu+1),a}}(X^{\mu-\nu-1}) - Q_{\mu,\nu}(a, b)$ is log-concave on $(0, \infty)$ for all $\mu \geq 1$, $\nu \geq 1/2$.*

### B. The log-concavity of the Nuttall $Q-$function with respect to the order $\mu$ and $\nu$

In this subsection, we consider the log-concavity of the Nuttall $Q-$function with respect to $\mu$ for two different cases of the order $\mu$ and $\nu$: (1) $\mu - \nu$ is fixed, (2) $\nu$ is fixed. The derived results for these two cases are quite different.

We first study the log-concavity of the Nuttall $Q-$function when $\mu - \nu$ is fixed. Consider the integral of $x^{(\mu-\nu-1)/2} f_{\chi^2_{2(\nu+1),a}}(x)$ on an interval $(c,d) \subseteq (0, \infty)$, the following result is true.

**Lemma 7.** *The function $\nu \mapsto \int_c^d x^{(\mu-\nu-1)/2} f_{\chi^2_{2(\nu+1),a}}(x) dx$ is log-concave on $[0, \infty)$ for fixed $\mu - \nu \geq 1$, $(c,d) \subseteq (0, \infty)$ and $a > 0$.*

The proof of Lemma 7 is given in Appendix E. By an immediate application of Lemma 7, we obtain the following result.

**Theorem 6.** *Let $a > 0$ and $\mu - \nu \geq 1$ be fixed. Then the following assertions about the normalized Nuttall Q-function, $\mathcal{Q}_{\mu,\nu}(a,b)$, are true:*
(a) *The function $\nu \mapsto \mathcal{Q}_{\mu,\nu}(a,b)$ is log-concave on $[0, \infty)$ for all $b \geq 0$.*

(b) *The function $\nu \mapsto E_{\chi^2_{2(\nu+1),a}}(X^{\mu-\nu-1}) - \mathcal{Q}_{\mu,\nu}(a,b)$ is log-concave on $[0,\infty)$ for all $b > 0$.*

*Proof:* Substituting $a$ with $a^2$, $c$ with $b^2$ and $d$ with $\infty$ in Lemma 7, we obtain the result of part (a). Similarly, substituting $a$ with $a^2$, $c$ with 0, and $d$ with $b^2$ in Lemma 7, part (b) is also proved. ∎

Since $a^\nu$ is log-linear in $\nu$, we obtain a similar result for the standard Nuttall $Q-$function.

**Corollary 2.** *Let $a > 0$ and $\mu - \nu \geq 1$ be fixed. Then the following assertions about the standard Nuttall $Q-$function, $Q_{\mu,\nu}(a,b)$, are true:*

(a) *The function $\nu \mapsto Q_{\mu,\nu}(a,b)$ is log-concave on $[0,\infty)$ for all $b \geq 0$.*
(b) *The function $\nu \mapsto a^\nu E_{\chi^2_{2(\nu+1),a}}(X^{\mu-\nu-1}) - Q_{\mu,\nu}(a,b)$ is log-concave on $[0,\infty)$ for all $b > 0$.*

When $\nu \geq 0$ is fixed, we may expect the Nuttall $Q-$function to be log-concave in $\mu$. However, the Nuttall $Q-$function is actually log-convex in $\mu$ in this case.

**Theorem 7.** *Let $\nu \geq 0$ be fixed and $a > 0$. Then the following assertions about the normalized and standard Nuttall $Q-$function, i.e. $\mathcal{Q}_{\mu,\nu}(a,b)$ and $Q_{\mu,\nu}(a,b)$, are true:*

(a) *The functions $\mu \mapsto \mathcal{Q}_{\mu,\nu}(a,b)$ and $\mu \mapsto Q_{\mu,\nu}(a,b)$ are log-convex on $[0,\infty)$ for $b \geq 0$.*
(b) *The functions $\mu \mapsto E_{\chi^2_{2(\nu+1),a}}(X^{\mu-\nu-1}) - \mathcal{Q}_{\mu,\nu}(a,b)$ and $\mu \mapsto a^\nu E_{\chi^2_{2(\nu+1),a}}(X^{\mu-\nu-1}) - Q_{\mu,\nu}(a,b)$ are log-convex on $[0,\infty)$ for $b > 0$.*

*Proof:* (a) We know that the integrand of (3) is log-linear in $\mu$, and hence is log-convex in $\mu$ for $\nu, \mu \geq 0$ and $a > 0$. Moreover, if function $f(x,y)$ is log-convex in $x$ for each $y \in C$, then

$$g(x) = \int_C f(x,y)dy \qquad (36)$$

is also log-convex [47, p. 106]. Therefore, $\mu \mapsto \mathcal{Q}_{\mu,\nu}(a,b)$ is log-convex on $[0,\infty)$. By multiplying $a^\nu$, we obtain that $\mu \mapsto Q_{\mu,\nu}(a,b)$ is also log-convex on $[0,\infty)$ and part (a) is proved.

(b) It is known that

$$E_{\chi^2_{2(\nu+1),a}}(X^{\mu-\nu-1}) - \mathcal{Q}_{\mu,\nu}(a,b) = \int_0^b \frac{t^\mu}{a^\nu} e^{-\frac{t^2+a^2}{2}} I_\nu(at) dt, \qquad (37)$$

which has the same integrand with (3), therefore the results of part (b) can be proved with the same argument. ∎

### C. The log-concavity of the Nuttall $Q-$function with respect to $a$

For the first step, we still consider the integral of $x^{(\mu-\nu-1)/2} f_{\chi^2_{2(\nu+1),a}}(x)$ on an interval $(c,d) \subseteq (0,\infty)$, like for the generalized Marcum $Q-$function. The following result is true.

**Lemma 8.** *The function $a \mapsto \int_c^d x^{(\mu-\nu-1)/2} f_{\chi^2_{2(\nu+1),a}}(x) dx$ is log-concave on $(0,\infty)$ for fixed $\mu - \nu \geq 1$, $(c,d) \subseteq (0,\infty)$ and $\nu \geq 0$.*

The proof of Lemma 8 is given in Appendix F. It is actually quite similar with that of Lemma 6. Substituting the integral limits in Lemma 8, we obtain the following result.

**Theorem 8.** *Let $\mu, \nu \geq 0$ and $\mu - \nu \geq 1$ be fixed. Then the following assertions about the normalized Nuttall $Q$-function, $\mathcal{Q}_{\mu,\nu}(a,b)$, are true:*

(a) *The function $a \mapsto \mathcal{Q}_{\mu,\nu}(\sqrt{a},b)$ is log-concave on $(0,\infty)$ for $b \geq 0$.*
(b) *The function $a \mapsto E_{\chi^2_{2(\nu+1),a}}(X^{(\mu-\nu-1)/2}) - \mathcal{Q}_{\mu,\nu}(\sqrt{a},b)$ is log-concave on $(0,\infty)$ for $b > 0$.*

*Proof:* (a) & (b) Tending with $c$ to $b^2$, and with $d$ to $\infty$ in Lemma 8, part (a) is proved. Moveover, substituting $c$ with 0 and $d$ with $b^2$ in Lemma 8, part (b) is also proved. ∎

These results can be easily generalized to the standard Nuttall $Q-$function.

**Corollary 3.** *Let $\mu, \nu \geq 0$ and $\mu - \nu \geq 1$ be fixed. Then the following assertions about the standard Nuttall $Q$-function, $Q_{\mu,\nu}(a,b)$, are true:*

(a) *The function $a \mapsto Q_{\mu,\nu}(\sqrt{a},b)$ is log-concave on $(0,\infty)$ for $b \geq 0$.*
(b) *The function $a \mapsto a^{\nu/2} E_{\chi^2_{2(\nu+1),a}}(X^{(\mu-\nu-1)/2}) - Q_{\mu,\nu}(\sqrt{a},b)$ is log-concave on $(0,\infty)$ for $b > 0$.*

## VI. BOUNDS OF THE GENERALIZED MARCUM AND NUTTALL $Q-$FUNCTIONS AND THEIR TIGHTNESS

### A. Closed-form expressions of the generalized Marcum and Nuttall $Q-$functions with special order

Recently, a closed-form expression of the generalized Marcum $Q-$function, $Q_\nu(a,b)$, was proposed for the case when $\nu$ is an odd multiple of 0.5, given by [32, eq. (11)]

$$\begin{aligned} &Q_\nu(a,b) \\ &= \frac{1}{2}\text{erfc}\left(\frac{b+a}{\sqrt{2}}\right) + \frac{1}{2}\text{erfc}\left(\frac{b-a}{\sqrt{2}}\right) \\ &+ \frac{1}{a\sqrt{2\pi}} \sum_{k=0}^{\nu-1.5} \frac{b^{2k}}{2^k} \sum_{q=0}^{k} \frac{(-1)^q (2q)!}{(k-q)!q!} \\ &\times \sum_{i=0}^{2q} \frac{1}{(ab)^{2q-i} i!} \left[(-1)^i e^{-\frac{(b-a)^2}{2}} - e^{-\frac{(b+a)^2}{2}}\right], \\ &\hspace{4cm} a > 0, b \geq 0. \end{aligned} \qquad (38)$$

where $\text{erfc}(\cdot)$ is the complementary error function [49, eq. (7.1.2)]. For the case $a = 0$, the value of the generalized Marcum $Q-$function is [32, eq. (12)]

$$Q_\nu(0,b) = \text{erfc}\left(\frac{b}{\sqrt{2}}\right) + \frac{e^{-\frac{b^2}{2}}}{\sqrt{2\pi}} \sum_{k=0}^{\nu-1.5} \frac{b^{2k+1}}{2^{k-1}} \sum_{q=0}^{k} \frac{(-1)^q}{(k-q)!q!(2q+1)}. \qquad (39)$$

where $\nu$ is an odd multiple of 0.5.

More compact closed-form expressions for the case $a > 0$ were derived in [21] and [33], based on the recursion formula of $Q_\nu(a,b)$ [4, eq. (4.34)], given by

$$Q_\nu(a,b) = Q(b+a) + Q(b-a)$$
$$+ a\sqrt{\frac{2}{\pi}}e^{-\frac{(a+b)^2}{2}}\sum_{n=1}^{\nu-0.5}(-2a^2)^{-n}$$
$$\times \sum_{k=0}^{n-1}\frac{(n-k)_{n-1}}{k!}(2ab)^k\left[1-(-1)^k e^{2ab}\right],$$
$$\nu + 0.5 \in \mathbb{N}, a > 0, \quad (40)$$

where $(\cdot)_\cdot$ is the Pochhammer's symbol [49, eq. (6.1.22)], $Q(x) = \frac{1}{2}\text{erfc}\left(\frac{x}{\sqrt{2}}\right)$ and

$$Q_\nu(a,b)$$
$$= \frac{1}{2}\text{erfc}\left(\frac{b+a}{\sqrt{2}}\right) + \frac{1}{2}\text{erfc}\left(\frac{b-a}{\sqrt{2}}\right)$$
$$+ \frac{1}{\sqrt{2\pi ab}}e^{-\frac{a^2+b^2}{2}}\sum_{k=0}^{\nu-1.5}\left(\frac{b}{a}\right)^{k+0.5}$$
$$\times \left\{\sum_{r=0}^{k}\frac{(k+r)!}{r!(k-r)!(2ab)^r}\left[(-1)^r e^{ab} + (-1)^{k+1}e^{-ab}\right]\right\},$$
$$a > 0, b \geq 0. \quad (41)$$

We note that the main difference of (40) and (41) is that different formulas of the modified Bessel function of the first kind were used during the course of their derivations. For the case $a = 0$, the generalized Marcum $Q$-function can be also expressed as [4, eq. (4.71)]

$$Q_\nu(0,b) = \frac{\Gamma\left(\nu,\frac{b^2}{2}\right)}{\Gamma(\nu)}, \quad (42)$$

where $\Gamma(\cdot,\cdot)$ is the upper incomplete gamma function [49, eq. (6.5.3)].

If $\nu$ is a positive integer, (42) reduces to [4, eq. (4.73)]

$$Q_\nu(0,b) = e^{-\frac{b^2}{2}}\sum_{m=0}^{\nu-1}\frac{(b^2/2)^m}{m!}. \quad (43)$$

Moreover, we can derive a novel formula for $Q_\nu(0,b)$, when $\nu$ is an odd multiple of 0.5. It is known that [49, eq. 8.356.2]

$$\Gamma(a+1, x) = a\Gamma(a,x) + x^a e^{-x}, \quad (44)$$

and [49, eq. 8.339.2]

$$\Gamma\left(n + \frac{1}{2}\right) = \frac{\sqrt{\pi}}{2^n}(2n-1)!!, \quad (45)$$

where $n$ is a positive integer and $(2n+1)!! = 1 \cdot 3 \ldots (2n+1)$ [63, p. xliii]. Therefore, we can get the recursion formula of $Q_\nu(0,b)$ for $\nu$ odd multiple of 0.5 after some manipulations

$$Q_{\nu+1}(0,b) = Q_\nu(0,b) + e^{-\frac{b^2}{2}}\sqrt{\frac{2}{\pi}}\frac{b^{2\nu}}{(2\nu)!!}. \quad (46)$$

From the integral of (20), we have

$$Q_{0.5}(0,b) = \text{erfc}\left(\frac{b}{\sqrt{2}}\right) = 2Q(b), \quad (47)$$

where $b \geq 0$.

Therefore, we obtain a new closed-form expression of $Q_\nu(0,b)$ for $\nu$ odd multiple of 0.5

$$Q_\nu(0,b) = \text{erfc}\left(\frac{b}{\sqrt{2}}\right) + e^{-\frac{b^2}{2}}\sqrt{\frac{2}{\pi}}\sum_{k=0}^{\nu-1.5}\frac{b^{2k+1}}{(2k+1)!!}, \quad (48)$$

which is more compact than (39).

Similar result was proposed for the standard Nuttall $Q$-function in [21, Theorem 1], given by

$$Q_{\mu,\nu}(a,b)$$
$$= \frac{(-1)^n(2a)^{-n+\frac{1}{2}}}{\sqrt{\pi}}\sum_{k=0}^{n-1}\frac{(n-k)_{n-1}(2a)^k}{k!}\mathcal{I}_{m,n}^k(a,b), \quad (49)$$

where $a > 0, b \geq 0, \mu \geq \nu, m = \mu + 0.5 \in \mathbb{N}, n = \nu + 0.5 \in \mathbb{N}$ and the term $\mathcal{I}_{m,n}^k(a,b)$ is given by

$$\mathcal{I}_{m,n}^k(a,b)$$
$$= (-1)^{k+1}\sum_{l=0}^{m-n+k}\binom{m-n+k}{l}2^{\frac{l-1}{2}}a^{m-n+k-l}$$
$$\left[\Gamma\left(\frac{l+1}{2}\right) + (-1)^{m-n-l-1}\Gamma\left(\frac{l+1}{2},\frac{(b+a)^2}{2}\right)\right.$$
$$\left. - [\text{sgn}(b-a)]^{l+1}\gamma\left(\frac{l+1}{2},\frac{(b-a)^2}{2}\right)\right], \quad (50)$$

where $\gamma(z,x) = \Gamma(z) - \Gamma(z,x)$ is the lower incomplete gamma function [49, eq. (6.5.2)], $\binom{\cdot}{\cdot}$ is the binomial coefficient [63, p. xliii] and $\text{sgn}(\cdot)$ is the signum function. We note that the term $\mathcal{I}_{m,n}^k(a,b)$ can be further expressed by $\text{erfc}(\cdot)$ and exponential function with the help of (42), (43) and (48).

The normalized Nuttall $Q$-function $\mathcal{Q}_{\mu,\nu}(a,b)$, for $m = \mu + 0.5 \in \mathbb{N}, n = \nu + 0.5 \in \mathbb{N}$, can be evaluated by [21, Corollary 1]

$$\mathcal{Q}_{\mu,\nu}(a,b)$$
$$= \frac{(-1)^n 2^{-n+\frac{1}{2}}}{\sqrt{\pi}a^{2n-1}}\sum_{k=0}^{n-1}\frac{(n-k)_{n-1}(2a)^k}{k!}\mathcal{I}_{m,n}^k(a,b), \quad (51)$$

where the term $\mathcal{I}_{m,n}^k(a,b)$ is given by (50).

### B. Upper and lower bounds for the generalized Marcum and Nuttall $Q$-functions

In [21] and [32], upper and lower bounds for the generalized Marcum and Nuttall $Q$-functions were proposed by using the monotonicity of these functions. We find that the log-concavity of the generalized Marcum and Nuttall $Q$-functions can be used to establish even tighter bounds.

*1) Upper and lower bounds for the generalized Marcum $Q$-function with real order $\nu$:* Although a lot of previous works only considered the generalized Marcum $Q$-function $Q_\nu(a,b)$ of integer order, the bounds for $Q_\nu(a,b)$ of non-integer order are desirable. Let $\lfloor x \rfloor$ be the maximal integer less than or equal to $x$. Then, $\nu_1 = \lfloor \nu + 0.5 \rfloor + 0.5$ is the minimal order that is larger than $\nu$ and also an odd multiple of 0.5, $\nu_2 = \lfloor \nu - 0.5 \rfloor + 0.5$ is the maximal order that is less than or equal to $\nu$ and is an odd multiple of 0.5. Since we have closed-form formula of $Q_\nu(a,b)$ for $\nu$ that is an odd multiple of 0.5. The log-concavity of $\nu \mapsto Q_\nu(a,b)$ on $[1,\infty)$, given in part (b) of Theorem 3, implies one lower bound of $Q_\nu(a,b)$

$$Q_\nu(a,b) \geq Q_{\nu-LB1}(a,b) \\ = Q_{\nu_1}(a,b)^{\nu-\nu_2} Q_{\nu_2}(a,b)^{\nu_1-\nu}, \ \nu \geq 1.5. \quad (52)$$

and also two more inequalities

$$Q_{\nu_1}(a,b) \\ \geq Q_\nu(a,b)^{\frac{1}{\nu_1-\nu+1}} Q_{\nu_1+1}(a,b)^{\frac{\nu_1-\nu}{\nu_1-\nu+1}}, \ \nu \geq 1, \quad (53)$$
$$Q_{\nu_2}(a,b) \\ \geq Q_\nu(a,b)^{\frac{1}{\nu-\nu_2+1}} Q_{\nu_2-1}(a,b)^{\frac{\nu-\nu_2}{\nu-\nu_2+1}}, \ \nu \geq 2.5. \quad (54)$$

After some simple algebraic manipulations, two upper bounds for $Q_\nu(a,b)$ are obtained, given by

$$Q_\nu(a,b) \leq Q_{\nu-UB1}(a,b) \\ = Q_{\nu_1}(a,b)^{\nu_1-\nu+1}/Q_{\nu_1+1}(a,b)^{\nu_1-\nu}, \ \nu \geq 1, \quad (55)$$

and

$$Q_\nu(a,b) \leq Q_{\nu-UB2}(a,b) \\ = Q_{\nu_2}(a,b)^{\nu-\nu_2+1}/Q_{\nu_2-1}(a,b)^{\nu-\nu_2}, \ \nu \geq 2.5. \quad (56)$$

Recall that in Theorem 1, we obtained that the function $\nu \mapsto Q_\nu(a,b)$ is strictly increasing for $\nu \in (0,\infty)$. Using this result, we can easily obtain that

$$Q_{\nu_2}(a,b) \leq Q_{\nu-LB1}(a,b) \leq Q_\nu(a,b) \\ \leq Q_{\nu-UB1}(a,b) < Q_{\nu_1}(a,b), \quad (57)$$

where $a \geq 0$, $b > 0$ and $\nu \geq 1.5$. This means that our bounds $Q_{\nu-LB1}(a,b)$ and $Q_{\nu-UB1}(a,b)$ are tighter than $Q_{\nu_2}(a,b)$ and $Q_{\nu_1}(a,b)$ proposed in [21], [32].

We can also use the log-concavity of the function $\nu \mapsto Q_\nu(a,b) - Q_\nu(0,b)$ on $(0,\infty)$, given in part (c) of Theorem 3, to get new bounds of $Q_\nu(a,b)$. We have the following inequalities

$$Q_\nu(a,b) - Q_\nu(0,b) \\ \geq [Q_{\nu_1}(a,b) - Q_{\nu_1}(0,b)]^{\nu-\nu_2} \\ \times [Q_{\nu_2}(a,b) - Q_{\nu_2}(0,b)]^{\nu_1-\nu}, \ \nu \geq 0.5, \quad (58)$$
$$Q_{\nu_1}(a,b) - Q_{\nu_1}(0,b) \\ \geq [Q_\nu(a,b) - Q_\nu(0,b)]^{\frac{1}{\nu_1-\nu+1}} \\ \times [Q_{\nu_1+1}(a,b) - Q_{\nu_1+1}(0,b)]^{\frac{\nu_1-\nu}{\nu_1-\nu+1}}, \ \nu > 0, \quad (59)$$
$$Q_{\nu_2}(a,b) - Q_{\nu_2}(0,b) \\ \geq [Q_\nu(a,b) - Q_\nu(0,b)]^{\frac{1}{\nu-\nu_2+1}} \\ \times [Q_{\nu_2-1}(a,b) - Q_{\nu_2-1}(0,b)]^{\frac{\nu-\nu_2}{\nu-\nu_2+1}}, \ \nu \geq 1.5. \quad (60)$$

Therefore, we obtain another lower bound of $Q_\nu(a,b)$, given by

$$Q_\nu(a,b) \geq Q_{\nu-LB2}(a,b) \\ = Q_\nu(0,b) + [Q_{\nu_1}(a,b) - Q_{\nu_1}(0,b)]^{\nu-\nu_2} \\ \times [Q_{\nu_2}(a,b) - Q_{\nu_2}(0,b)]^{\nu_1-\nu}, \ \nu \geq 0.5, \quad (61)$$

and two upper bounds

$$Q_\nu(a,b) \leq Q_{\nu-UB3}(a,b) \\ = Q_\nu(0,b) + [Q_{\nu_1}(a,b) - Q_{\nu_1}(0,b)]^{\nu_1-\nu+1} \\ /[Q_{\nu_1+1}(a,b) - Q_{\nu_1+1}(0,b)]^{\nu_1-\nu}, \ \nu > 0, \quad (62)$$

and

$$Q_\nu(a,b) \leq Q_{\nu-UB4}(a,b) \\ = Q_\nu(0,b) + [Q_{\nu_2}(a,b) - Q_{\nu_2}(0,b)]^{\nu-\nu_2+1} \\ /[Q_{\nu_2-1}(a,b) - Q_{\nu_2-1}(0,b)]^{\nu-\nu_2}, \ \nu \geq 1.5. \quad (63)$$

We note that the log-concavity of the function $\nu \mapsto 1 - Q_\nu(a,b)$ can be also used to generate bounds of $Q_\nu(a,b)$, but the derived bounds are not reliable for large $b$.

*2) Upper and lower bounds for the Nuttall $Q$–function when $\mu - \nu \geq 1$ is an integer:* Let $\mu_1 = \lfloor \mu + 0.5 \rfloor + 0.5$ is the minimal order that is larger than $\mu$ and also an odd multiple of 0.5, $\mu_2 = \lfloor \mu - 0.5 \rfloor + 0.5$ is the maximal order that is less than or equal to $\mu$ and is an odd multiple of 0.5. In Theorem 6, we obtained that the function $\nu \mapsto \mathcal{Q}_{\mu,\nu}(a,b)$ is log-concave on $[0,\infty)$ for $\mu - \nu \geq 1$ fixed. Therefore, we can get a lower bound for the normalized Nuttall $Q$–function when $\mu - \nu \geq 1$ is an integer, given by

$$\mathcal{Q}_{\mu,\nu}(a,b) \geq \mathcal{Q}_{\mu,\nu-LB}(a,b) \\ = \mathcal{Q}_{\mu_1,\nu_1}(a,b)^{\nu-\nu_2} \mathcal{Q}_{\mu_2,\nu_2}(a,b)^{\nu_1-\nu}, \ \nu \geq 0.5. \quad (64)$$

Moreover, we can get two upper bounds for $\mathcal{Q}_{\mu,\nu}(a,b)$ with $\mu - \nu \geq 1$ an integer after some simple manipulations, given by

$$\mathcal{Q}_{\mu,\nu}(a,b) \leq \mathcal{Q}_{\mu,\nu-UB1}(a,b) \\ = \mathcal{Q}_{\mu_1,\nu_1}(a,b)^{\nu_1-\nu+1}/\mathcal{Q}_{\mu_1+1,\nu_1+1}(a,b)^{\nu_1-\nu}, \ \nu \geq 0, \quad (65)$$

and

$$\mathcal{Q}_{\mu,\nu}(a,b) \leq \mathcal{Q}_{\mu,\nu-UB2}(a,b) \\ = \mathcal{Q}_{\mu_2,\nu_2}(a,b)^{\nu-\nu_2+1}/\mathcal{Q}_{\mu_2-1,\nu_2-1}(a,b)^{\nu-\nu_2}, \ \nu \geq 1.5. \quad (66)$$

In Theorem 1, we proved that the function $\nu \mapsto \mathcal{Q}_{\mu+\nu+1,\nu}(a,b)$ is strictly increasing. Therefore, we have that

$$\mathcal{Q}_{\mu_2,\nu_2}(a,b) \leq \mathcal{Q}_{\mu,\nu-LB}(a,b) \leq \mathcal{Q}_{\mu,\nu}(a,b) \\ \leq \mathcal{Q}_{\mu,\nu-UB1}(a,b) < \mathcal{Q}_{\mu_1,\nu_1}(a,b), \quad (67)$$

where $a,b > 0$, $\nu \geq 0.5$ and $\mu - \nu \geq 1$ is an integer. This means the proposed bounds $\mathcal{Q}_{\mu,\nu-LB}(a,b)$ and $\mathcal{Q}_{\mu,\nu-UB1}(a,b)$ are tighter than those given in [21].

These results can be also generalized to the standard Nuttall $Q$–function, $Q_{\mu,\nu}(a,b)$ with $\mu - \nu \geq 1$ an integer, with one lower bound given by

$$Q_{\mu,\nu}(a,b) \geq Q_{\mu,\nu-LB}(a,b) \\ = Q_{\mu_1,\nu_1}(a,b)^{\nu-\nu_2} Q_{\mu_2,\nu_2}(a,b)^{\nu_1-\nu}, \ \nu \geq 0.5, \quad (68)$$

and two upper bounds

$$Q_{\mu,\nu}(a,b) \leq Q_{\mu,\nu-UB1}(a,b) \\ = Q_{\mu_1,\nu_1}(a,b)^{\nu_1-\nu+1}/Q_{\mu_1+1,\nu_1+1}(a,b)^{\nu_1-\nu}, \ \nu \geq 0, \quad (69)$$

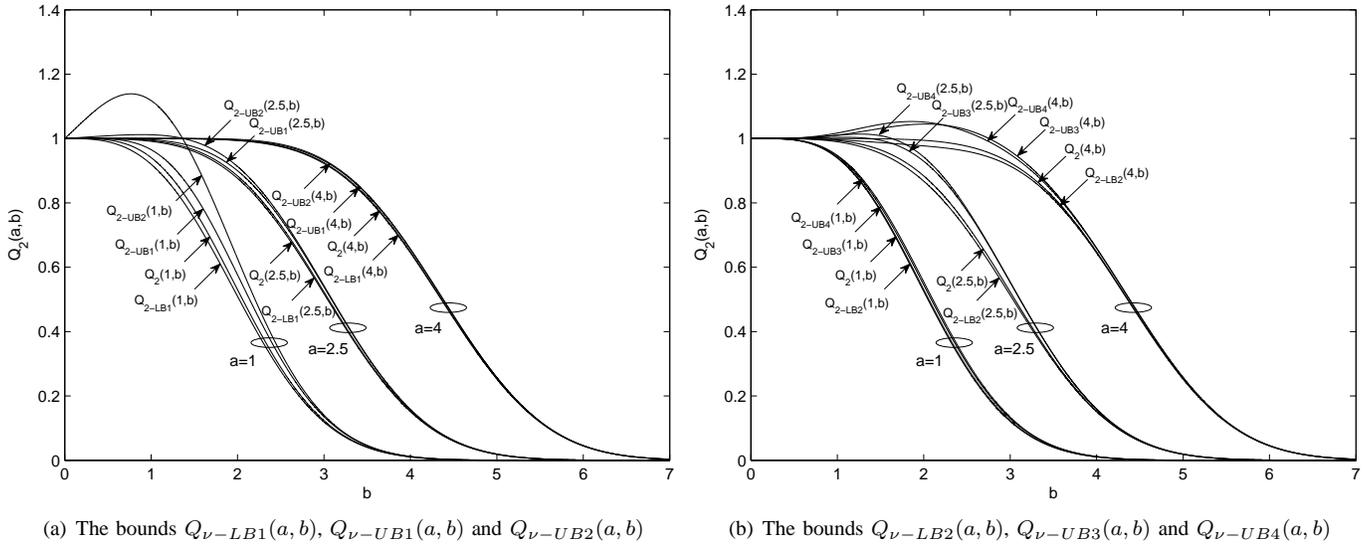

Fig. 4. Numerical results for $Q_\nu(a,b)$ and the proposed bounds versus $b$ for $a \in \{1, 2.5, 4\}$ and $\nu = 2$.

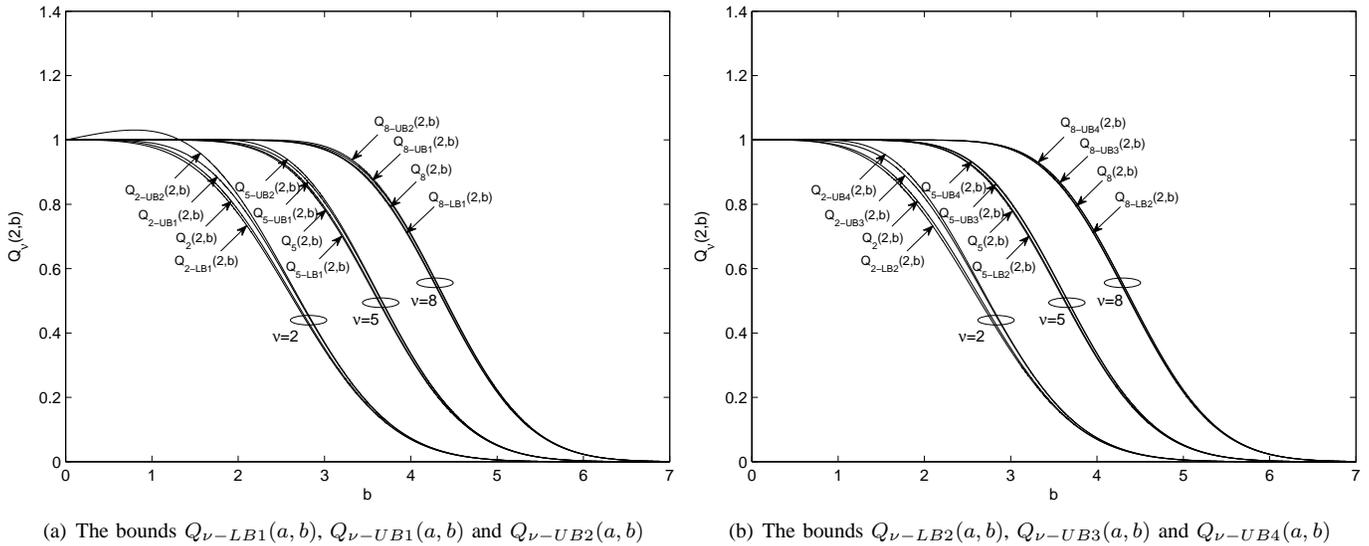

Fig. 5. Numerical results for $Q_\nu(a,b)$ and the proposed bounds versus $b$ for $\nu \in \{2, 5, 8\}$ and $a = 2$.

and

$$Q_{\mu,\nu}(a,b) \leq Q_{\mu,\nu-UB2}(a,b)$$
$$= Q_{\mu_2,\nu_2}(a,b)^{\nu-\nu_2+1}/Q_{\mu_2-1,\nu_2-1}(a,b)^{\nu-\nu_2}, \ \nu \geq 1.5. \quad (70)$$

We note that these bounds for the standard Nuttall $Q$−function $Q_{\mu,\nu}(a,b)$ are tighter than those given in [21] when $a \geq 1$ by means of (67). Moreover, these bounds hold true when $0 < a < 1$, while the bounds given in [21] do not.

### C. The tightness of the bounds

*1) The bounds of the generalized Marcum Q−function:* For the bounds of the generalized Marcum $Q$−function $Q_\nu(a,b)$, we first consider their tightness with respect to different values of the parameters. Fig. 4 shows the bounds of $Q_\nu(a,b)$ with different values of $a$, i.e. $a \in \{1, 2.5, 4\}$, when $\nu = 2$; Fig. 5 shows the results for different values of $\nu$, i.e. $\nu \in$ $\{2, 5, 8\}$, when $a = 2$. We observe that the tightness of the bounds $Q_{\nu-LB1}(a,b)$, $Q_{\nu-UB1}(a,b)$ and $Q_{\nu-UB2}(a,b)$ improves as either $a$ or $\nu$ increases, and the tightness of the bounds $Q_{\nu-LB2}(a,b)$, $Q_{\nu-UB3}(a,b)$ and $Q_{\nu-UB4}(a,b)$ improves as $\nu$ increases, but worsens as $a$ increases. Therefore, $Q_{\nu-LB1}(a,b)$, $Q_{\nu-UB1}(a,b)$ and $Q_{\nu-UB2}(a,b)$ are more suitable for large values of $a$, while $Q_{\nu-LB2}(a,b)$, $Q_{\nu-UB3}(a,b)$ and $Q_{\nu-UB4}(a,b)$ are proper for relative small $a$. However, since $Q_\nu(0,b)$ becomes very small as $b$ grows, these two groups of bounds tends to be equal for large $b$.

The numerical results of our bounds $Q_{\nu-LB2}(a,b)$, $Q_{\nu-UB3}(a,b)$ and $Q_{\nu-UB4}(a,b)$ with non-integer order $\nu$ are shown in Fig. 6. We can see that our new bounds are tighter than the bounds proposed in [21]. It is interesting to see that the upper bound $Q_{\nu-UB3}(a,b)$ is tighter than $Q_{\nu-UB4}(a,b)$ when $\nu = 5.1$, but $Q_{\nu-UB4}(a,b)$ can be tighter than $Q_{\nu-UB3}(a,b)$ when $\nu = 1.8$. This is because $\nu_1 - \nu = 0.4 < \nu - \nu_2 = 0.6$ when $\nu = 5.1$, and the inequality in (59) can be tighter than that in (60). When $\nu = 1.8$, $\nu_1 - \nu = 0.7 > \nu - \nu_2 = 0.3$ and the inequality in (59) is looser than that in (60). Similar results can be also found for the bounds $Q_{\nu-UB1}(a,b)$ and $Q_{\nu-UB2}(a,b)$. Therefore, we can choose the tighter upper and lower bounds proposed in this paper according to the value of $a$ and the decimal value of $\nu$.

Next, we compare the proposed bounds with other existing bounds when $\nu$ is integer. Since most of the existing bounds for the generalized Marcum $Q-$function of integer $\nu$ order are valid for only either $b > a$ or $b < a$, we choose to show the comparisons case by case.

For the case $b > a$, the existing lower bounds include LB1-AT in [35, the first line in eq. (18)][1], $Q_{m-0.5}(a,b)$ in [32, eq. (11) and (14)], GLBm1-KL in [38, eq. (6)], LB1-BS in [41, eq. (4) and (8)] and LB1-B in [42, eq. (8) and (15)]. The existing upper bounds include UB1-SA in [34, eq. (8)], UB1-AT in [35, eq.(17)], $Q_{m+0.5}(a,b)$ in [32, eq. (11) and (14)], GUBm1-KL in [38, eq. (5)], UB1-BS in [41, eq. (5) and (9)] and UB1-B in [42, eq. (9) and (16)]. Fig. 7 shows the results for the case $b > a = 1.5$ and $\nu = 2$ in a logarithmic scale. We choose relative small values of $\nu$ and $a$ in order to facilitate the recognition of our bounds from the exact value. Even in this case, our new bounds are shown to be much tighter than the other bounds in the literature. For larger values of $\nu$ and $a$, our numerical results show that our bounds are very tight.

For the case $b < a$, the existing lower bounds include LB2-SA in [34, eq. (12)], LB1-AT in [35, the first line in eq. (18)], LB2-AT in [35, eq. (20)], LB3-AT in [35, eq. (21)], $Q_{\nu-0.5}(a,b)$ in [32, eq. (11) and (14)], GLBm2-KL in [38, eq. (9)], LB2-BS in [41, eq. (11) and (17)] and LB2-B in [42, eq. (17) and (23)]. The existing upper bounds include $Q_{m+0.5}(a,b)$ in [32, eq. (11) and (14)], GUBm2-KL in [38, eq. (8)], UB2-BS in [41, eq. (11) and (17)] and UB2-B in [42, eq. (18) and (24)]. The numerical results for the case $b < a = 4$ and $\nu = 4$ are illuminated in Fig. 8. Our bounds $Q_{\nu-LB1}(a,b)$ and $Q_{\nu-UB1}(a,b)$ are much tighter than the other bounds in most of the cases, but the $Q_{\nu-UB2}(a,b)$ can be looser than the existing bounds.

Let the absolute relative error of a bound be $\varepsilon\% = 100\% \times \frac{|\text{bound}-Q_\nu(a,b)|}{Q_\nu(a,b)}$. If $\nu = 4$ and $a = 4$, the maximal absolute relative errors of our bounds $Q_{\nu-LB1}(a,b)$, $Q_{\nu-UB1}(a,b)$ and $Q_{\nu-UB2}(a,b)$ are 0.4437%, 1.3077% and 1.3799%, respectively, for all range of $b$. If $\nu = 6$ and $a = 6$, the the maximal absolute relative errors of our bounds reduce to 0.2157%, 0.6420% and 0.6581%, respectively. Our numerical results suggest that the absolute relative errors of the proposed bounds are less than 5% in most of the cases.

Moreover, the relative errors of all our bounds converge to 0 as $b \to \infty$. This can be explained simply by the asymptotic formula of the generalized Marcum $Q$-function when $b \to \infty$, given in (4). From (4), we can find that $Q_\nu(a,b)$ inclines to log-linear for very large $b$ and the inequalities (52), (53) and (54) tend to be equal. Hence, the bounds $Q_{\nu-LB1}(a,b)$, $Q_{\nu-UB1}(a,b)$ and $Q_{\nu-UB2}(a,b)$ converge to the exact value as $b \to \infty$. Since $Q_\nu(0,b) \to 0$ as $b \to \infty$, we obtain that the bounds $Q_{\nu-LB2}(a,b)$, $Q_{\nu-UB3}(a,b)$ and $Q_{\nu-UB4}(a,b)$ possess the same property.

---

[1]There is a mistake in the formula given in [35]. For a correct version, the readers are referred to equation (43).

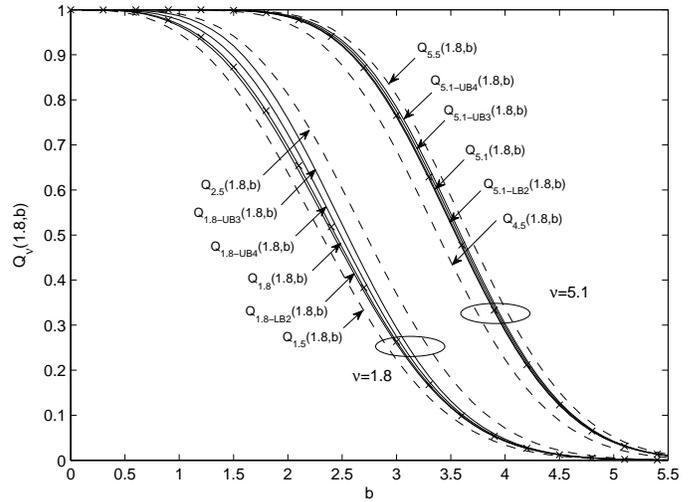

Fig. 6. Numerical results for the bounds of $Q_\nu(a,b)$ with non-integer order, where $\nu = 1.8, 5.1$ and $a = 1.8$. 'x': exact. Dashed line: previous bounds. Solid line: some of our new bounds including $Q_{\nu-LB2}(a,b)$, $Q_{\nu-UB3}(a,b)$ and $Q_{\nu-UB4}(a,b)$.

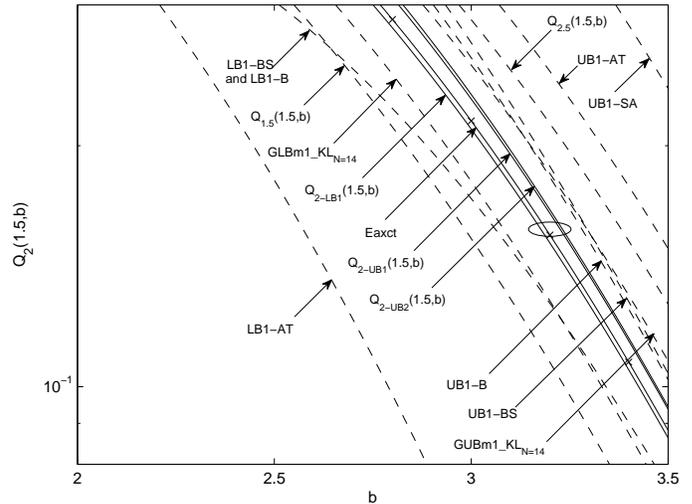

Fig. 7. Numerical results for $Q_\nu(a,b)$ and its upper and lower bounds versus $b$ for the case $b > a = 1.5$ and $\nu = 2$. 'x': exact. Dashed line: previous bounds. Solid line: some of our new bounds, including $Q_{\nu-LB1}(a,b)$, $Q_{\nu-UB1}(a,b)$ and $Q_{\nu-UB2}(a,b)$.

On the other hand, it was proved that the relative errors of the bounds $Q_{\nu-0.5}(a,b)$ and $Q_{\nu+0.5}(a,b)$ does not tend to zero as $b$ approaches infinity [41]. And the bounds proposed in [41], whose relative errors tend to zero when $b \to \infty$, hold true for only $b > a$. To the extent of the authors' knowledge, our bounds are the first bounds with such tightness on the whole region of $b > 0$, even in terms of relative errors.

*2) The bounds of the Nuttall $Q-$function:* We compare the proposed bounds of the normalized and standard Nuttall $Q-$function of the order $\mu, \nu \geq 0$ with other existing bounds.

The existing lower and upper bounds for the normalized

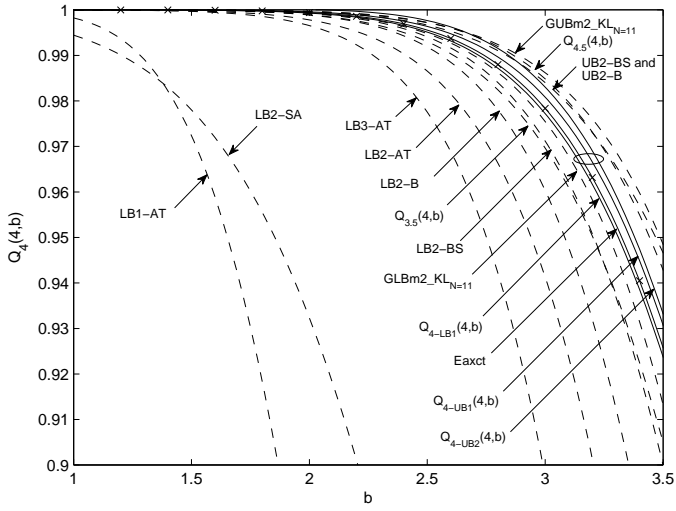

Fig. 8. Numerical results for $Q_\nu(a,b)$ and its upper and lower bounds versus $b$ for the case $b < a = 4$ and $\nu = 4$. 'x': exact. Dashed line: previous bounds. Solid line: some of our new bounds including $Q_{\nu-LB1}(a,b)$, $Q_{\nu-UB1}(a,b)$ and $Q_{\nu-UB2}(a,b)$.

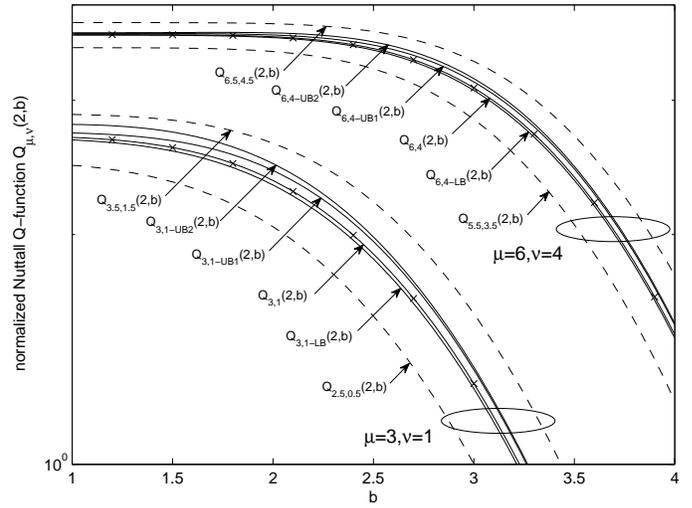

Fig. 10. Numerical results for the normalized Nuttall $Q-$function $\mathcal{Q}_{\mu,\nu}(a,b)$ and its bounds versus $b$ for different values of $\nu$ and $\mu$ with $\mu - \nu$ fixed, i.e. $\nu = 1, 4$ and $\mu - \nu = 2$, when $a = 2$. 'x': exact. Dashed line: previous bounds. Solid line: our new bounds including $\mathcal{Q}_{\mu,\nu-LB}(a,b)$, $\mathcal{Q}_{\mu,\nu-UB1}(a,b)$ and $\mathcal{Q}_{\mu,\nu-UB2}(a,b)$.

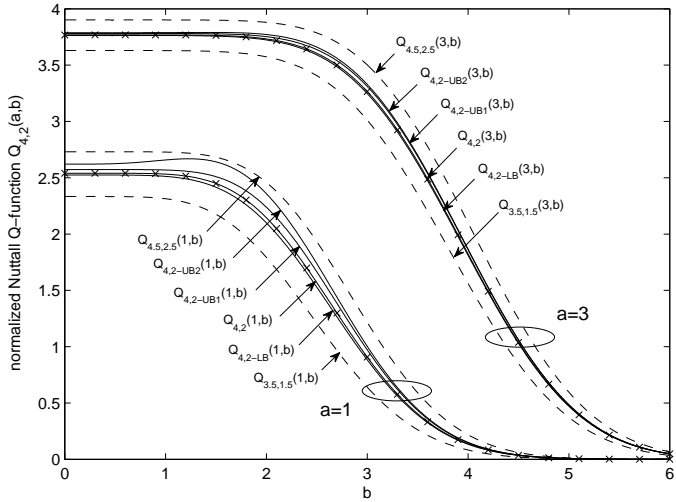

Fig. 9. Numerical results for the normalized Nuttall $Q-$function $\mathcal{Q}_{\mu,\nu}(a,b)$ and its bounds versus $b$ for different values of $a$, i.e. $a = 1, 3$, when $\mu = 4$ and $\nu = 2$. 'x': exact. Dashed line: previous bounds. Solid line: our new bounds including $\mathcal{Q}_{\mu,\nu-LB}(a,b)$, $\mathcal{Q}_{\mu,\nu-UB1}(a,b)$ and $\mathcal{Q}_{\mu,\nu-UB2}(a,b)$.

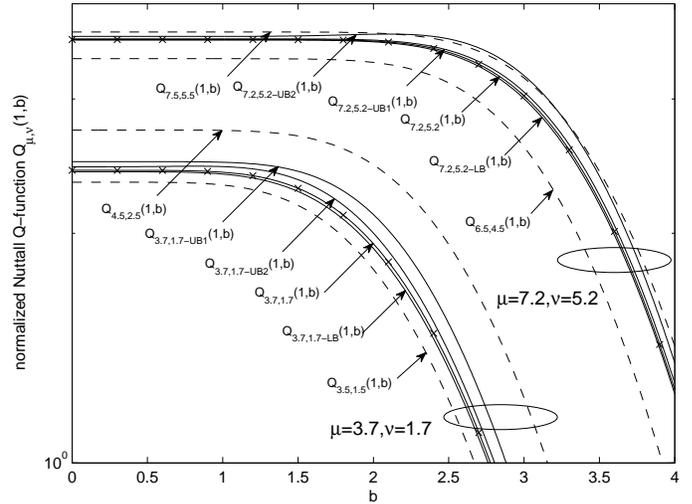

Fig. 11. Numerical results for the bounds of the normalized Nuttall $Q-$function $\mathcal{Q}_{\mu,\nu}(a,b)$ with non-integer order, where $\mu - \nu = 2$, $\nu = 1.7, 5.2$ and $a = 1$. 'x': exact. Dashed line: previous bounds. Solid line: our new bounds including $\mathcal{Q}_{\mu,\nu-LB}(a,b)$, $\mathcal{Q}_{\mu,\nu-UB1}(a,b)$ and $\mathcal{Q}_{\mu,\nu-UB2}(a,b)$.

Nuttall $Q-$function $\mathcal{Q}_{\mu,\nu}(a,b)$ are $\mathcal{Q}_{\lfloor\mu\rfloor_{0.5},\lfloor\nu\rfloor_{0.5}}(a,b)$ and $\mathcal{Q}_{\lceil\mu\rceil_{0.5},\lceil\nu\rceil_{0.5}}(a,b)$ [21, eq. 19], respectively, when $\mu \geq \nu + 1$ and $\nu \geq 1$. Fig. 9 shows the numerical results for different values of $a$, i.e. $a \in \{1, 3\}$, when $\mu = 4$ and $\nu = 2$; Fig. 10 shows the numerical results for different values of $\nu$ and $\mu$, i.e. $\nu \in \{1, 4\}$ and $\mu = \nu + 2$, when $a = 2$. We find that the tightness of our bounds $\mathcal{Q}_{\mu,\nu-LB}(a,b)$, $\mathcal{Q}_{\mu,\nu-UB1}(a,b)$ and $\mathcal{Q}_{\mu,\nu-UB2}(a,b)$ improves as either $a$ increases or $\nu$ increases with $\mu - \nu$ fixed. This result is expected for $\mathcal{Q}_{\mu,\nu}(a,b)$, since it holds true for the special case of the generalized Marcum $Q-$function $Q_\nu(a,b)$. As we have proved in (67), the proposed bounds $\mathcal{Q}_{\mu,\nu-LB}(a,b)$ and $\mathcal{Q}_{\mu,\nu-UB1}(a,b)$ are tighter than the bounds given in [21].

The numerical results for the case that $\nu$ is not an integer and $\mu - \nu \geq 1$ is an integer are shown in Fig. 11. We can see that the tighter ones of our new bounds are tighter than the bounds proposed in [21] when the orders $\mu$ and $\nu$ are not integer. Moreover, we find that the upper bound $Q_{\mu,\nu-UB1}(a,b)$ is tighter than $Q_{\mu,\nu-UB2}(a,b)$ when $\nu = 5.2$, while $Q_{\mu,\nu-UB2}(a,b)$ can be tighter than $Q_{\mu,\nu-UB1}(a,b)$ when $\nu = 1.7$. This observation is quite similar with the case of generalized Marcum $Q-$function. It means that we can choose the tighter upper bounds of the normalized Nuttall $Q-$function according to the decimal value of $\nu$.

The numerical results for the exact value and bounds of standard Nuttall $Q-$function are shown in Fig. 12. When $a > 1$, the proposed bounds are tighter than the bounds given in [21]. When $a < 1$, our proposed bounds hold true, while the bounds of [21] do not. The bound $Q_{\mu,\nu-UB1}(a,b)$ is also tighter than $Q_{\mu,\nu-UB2}(a,b)$, and we can use $Q_{\mu,\nu-UB1}(a,b)$ in the applications. The other properties for the bounds of normalized Nuttall $Q$−function maintain for the bounds of standard Nuttall $Q$−function in terms of relative errors.

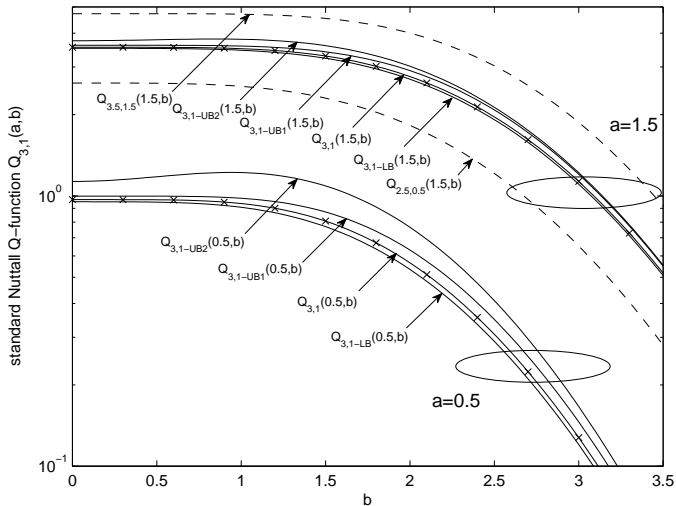

Fig. 12. Numerical results for the standard Nuttall $Q$−function $Q_{\mu,\nu}(a,b)$ and its bounds versus $b$ for for different values of $a$, i.e. $a = 0.5, 1.5$, when $\nu = 3$, $\mu = 1$. 'x': exact. Dashed line: previous bounds. Solid line: some of our new bounds including $Q_{\mu,\nu-LB}(a,b)$, $Q_{\mu,\nu-UB1}(a,b)$ and $Q_{\mu,\nu-UB2}(a,b)$.

Now we consider the tightness of our bounds for the Nuttall $Q$−function on the whole region of $b \in (0,\infty)$. For a bit larger parameters, when $\mu = 7$, $\nu = 4$ and $a = 4$, the maximal absolute relative errors of our bounds $Q_{\mu,\nu-LB}(a,b)$, $Q_{\mu,\nu-UB1}(a,b)$ and $Q_{\mu,\nu-UB2}(a,b)$ are 0.4134%, 1.2216% and 1.2802%, respectively, for all range of $b$. When $\mu = 9$, $\nu = 6$ and $a = 6$, the maximal absolute relative errors of our bounds reduce to 0.2079%, 0.6190% and 0.6334%, respectively. Our numerical results suggest that the absolute relative errors of the proposed bounds are less than 5% in most of the cases.

Using the same method given in [4, p. 81], we can get the asymptotic formula of the normalized Nuttall $Q$-function

$$\mathcal{Q}_{\mu,\nu}(a,b) \sim \frac{b^{\mu-0.5}}{a^{\nu+0.5}} Q(b-a). \qquad (71)$$

For the standard Nuttall $Q$-function, we have

$$Q_{\mu,\nu}(a,b) \sim \frac{b^{\mu-0.5}}{a^{0.5}} Q(b-a). \qquad (72)$$

It implies that the relative errors of our bounds of the normalized and standard Nuttall $Q$−functions also converge to 0 as $b \to \infty$. Moreover, using (71) it is quite simple to show that the relative errors of the bounds $\mathcal{Q}_{\lfloor\mu\rfloor_{0.5},\lfloor\nu\rfloor_{0.5}}(a,b)$ and $\mathcal{Q}_{\lceil\mu\rceil_{0.5},\lceil\nu\rceil_{0.5}}(a,b)$ do not tend to zero as $b$ approaches infinity. To the extent of the authors' knowledge, our bounds for the normalized and standard Nuttall $Q$−functions are the first bounds with such tightness in terms of relative errors on the whole region of $b > 0$.

## VII. APPLICATIONS

The proposed bounds for the generalized Marcum and Nuttall $Q$−functions have been shown to be quite tight in the previous section. They involve only exponential function and the erfc function, and therefore can be computed very efficiently. These results can be applied to the performance analysis of various wireless communication systems operating over fading channels. Some application examples of our proposed bounds are given as follows:

One interesting example occurs when bounding the outage probability of wireless communication systems which are both interference- and power-limited, where the Rayleigh/Nakagami faded desired signals are subject to independent and identically distributed (i.i.d.) Rician faded interferers [64, eq. (18) and (26)]. Another similar application is to bound the outage probability of maximal ratio combining (MRC) in the presence of independent but not necessarily identically distributed Rayleigh faded co-channel interference, when the received signal at every antenna experiences i.i.d. Ricean fading [22, eq. (21), (27) and (28)]. The techniques in these two examples apply directly in the outage probability analysis of cognitive radio (CR) system and wireless sensor networks (WSN) [67].

Our proposed results can also be applied to evaluate the average error probability of digital communication systems operating over slow-fading channels. In [25], the authors approximated the average error probability of MRC multichannel reception by using a piecewise polynomial approximation method. Their results require to compute the partial (truncated) general moment of generalized Rayleigh or Ricean distribution, which can be simply represented by generalized Marcum and Nuttall $Q$−functions. Actually, comparing with the infinite series formulation in [25, eq. (17)] for the case $s > 0$, one would prefer to use the result

$$\Upsilon_{g^2,j}(\tilde{\gamma}_i) - \Upsilon_{g^2,j}(\tilde{\gamma}_{i-1}) \\ = Q_{\frac{n}{2}+2j,\frac{n}{2}-1}\left(\frac{s}{\sigma},\frac{\sqrt{\tilde{\gamma}_{i-1}}}{\sigma}\right) - Q_{\frac{n}{2}+2j,\frac{n}{2}-1}\left(\frac{s}{\sigma},\frac{\sqrt{\tilde{\gamma}_i}}{\sigma}\right), \qquad (73)$$

where $g^2$ is subject to a distribution of $\sigma^2 \chi^2_{n,s^2/\sigma^2}$. We note that the order of generalized Rayleigh or Ricean distribution can be not integer. Therefore, the evaluation of non-integer order generalized Marcum and Nuttall $Q$−functions is needed.

Finally, our results can be used to extract the log-likelihood ratio for the decoding of turbo or low-density parity check (LDPC) codes for differential phase-shift keying (DPSK) signals [65].

## VIII. CONCLUSION

In this paper, we provided a comprehensive study of the monotonicity and log-concavity properties for the generalized Marcum and Nuttall $Q$−functions. Tight upper and lower bounds for the generalized Marcum and Nutall $Q$−functions have been obtained by using the log-concavity of these functions. If the bounds are chosen based on the values of the parameters, the proposed bounds are tighter than the existing bounds in the literature in most of the cases. Our proposed bounds are tight in terms of relative errors for all range of $b$. We have proved that the relative errors of our proposed bounds converge to 0 as $b \to \infty$. The numerical results show that the absolute relative errors of the proposed bounds are less than 5% in most of the cases. To the extent of the authors' knowledge, our bounds for the generalized Marcum and Nuttall $Q$-functions are the first bounds with such tightness in terms of relative errors on the whole region of $b \in (0, \infty)$. Some applications of the proposed theoretical results have been also provided. Further research directions include to prove the conjectures that were provided in Section IV.

## ACKNOWLEDGEMENT

The first author is very grateful to Professor Hao Zhang at Department of Electronic Engineering, Tsinghua University, for introducing him to the theory of the Marcum $Q$-function. The authors are grateful to the referees for their constructive comments and suggestions, which are very helpful for improving this paper.

## APPENDIX A
### PROOF OF LEMMA 1

It is known that $E_X(E_Y(Y|X=x)) = E_Y(Y)$ and $P(Y \in S) = E_Y(\mathcal{L}_S)$, where $\mathcal{L}_S$ is the indicator function of the set $S$. Therefore,

$$P(Y \in S) = E_X(P(Y \in S|X=x)). \quad (74)$$

Since $X$ and $Y$ are non-negative independent random variables, we have

$$\begin{aligned} F_{X+Y}(t) &= P(X+Y \le t) \\ &= E_X\left(P(x+Y \le t)|X=x\right) \\ &= \int_0^t F_Y(t-x)f_X(x)dx. \end{aligned} \quad (75)$$

The differential of $F_{X+Y}(t)$ is

$$\begin{aligned} & f_{X+Y}(t)dt \\ &= F_Y(0)f_X(t)dt + \int_0^t dF_Y(t-x)f_X(x)dx, \end{aligned} \quad (76)$$

where we have used the definition of Riemann-Stieltjes integrals. Since $F_Y(0) \ge 0$, we have

$$f_{X+Y}(t)dt \ge \int_0^t dF_Y(t-x)f_X(x)dx. \quad (77)$$

This in turn implies that

$$\begin{aligned} & \int_b^\infty g(t)f_{X+Y}(t)dt \\ &\ge \int_b^\infty g(t)\left(\int_0^t dF_Y(t-x)f_X(x)dx\right) \\ &= \int_0^\infty \left(\int_{\max\{x,b\}}^\infty g(t)dF_Y(t-x)\right)f_X(x)dx \\ &= \int_0^b \left(\int_b^\infty g(t)dF_Y(t-x)\right)f_X(x)dx \\ &\quad + \int_b^\infty \left(\int_x^\infty g(t)dF_Y(t-x)\right)f_X(x)dx \\ &= \int_0^b \left(\int_{b-x}^\infty g(x+y)dF_Y(y)\right)f_X(x)dx \\ &\quad + \int_b^\infty \left(\int_0^\infty g(x+y)dF_Y(y)\right)f_X(x)dx. \end{aligned} \quad (78)$$

Here, we note that the first term and the second term in above result are exactly the integrations on region $A$ and $B$ (see Fig. 1), respectively. Since $F_Y(0) < 1$, we know $dF_Y(y) > 0$ on a subset of $(0, \infty)$ with non-zero measure. This indicates the following inequality

$$\int_0^b \left(\int_{b-x}^\infty g(x+y)dF_Y(y)\right)f_X(x)dx > 0. \quad (79)$$

On the other hand, we know that

$$\begin{aligned} & \int_b^\infty \left(\int_0^\infty g(x+y)dF_Y(y)\right)f_X(x)dx \\ &\ge \int_b^\infty \left(\int_0^\infty dF_Y(y)\right)g(x)f_X(x)dx \\ &= \int_b^\infty g(x)f_X(x)dx. \end{aligned} \quad (80)$$

With this, the proof is complete.

## APPENDIX B
### PROOF OF THEOREM 3

(a) The proposed result is equivalent with the log-concavity of $\nu \mapsto Q_\nu(0, \sqrt{b})$. We have

$$\begin{aligned} Q_\nu\left(0, \sqrt{b}\right) &= \int_b^\infty f_{\chi^2_{2\nu}}(x)dx \\ &= \int_0^\infty f_{\chi^2_{2\nu}}(x)\mathcal{L}_{[b,\infty)}(x)dx, \end{aligned} \quad (81)$$

where $\mathcal{L}_{[b,\infty)}$ is the indicator function for the interval $[b, \infty)$. We can easily prove that $(x, y) \mapsto \mathcal{L}_{[b,\infty)}(x+y)$ is $SRR_2$ for $x, y > 0$ and $b \ge 0$ from the definition. Using the reproductive property of the central chi-square distribution one has

$$\begin{aligned} f_{\chi^2_{2(\nu_1+\nu_2)}}(x) &= f_{\chi^2_{2\nu_1}}(x) * f_{\chi^2_{2\nu_2}}(x) \\ &= \int_0^x f_{\chi^2_{2\nu_1}}(t)f_{\chi^2_{2\nu_2}}(x-t)dt. \end{aligned} \quad (82)$$

Applying parts (a) and (b) of Lemma 5 for the pdf of (central) chi-square distribution given in (10) we get that $(x, \nu) \mapsto f_{\chi^2_{2\nu}}(x)$ is $TP_2$ for $x, \nu > 0$. Then, from Lemma 4, we can easily obtain that $Q_{\nu_1+\nu_2}(0, \sqrt{b})$ is $SRR_2$ for $\nu_1 > 0$ and $\nu_2 > 0$. This implies that $\nu \mapsto Q_\nu(0, \sqrt{b})$ is log-concave for $\nu > 0$ due to part (d) of Lemma 5. Substituting $b$ with $b^2$, part (a) is proved.

(b) In view of (8) and (74), we get that

$$\begin{aligned} & Q_{\nu+1}\left(\sqrt{a}, \sqrt{b}\right) \\ &= P(\chi^2_{2\nu+2,a} \ge b) \\ &= P(\chi^2_{2\nu} + \chi^2_{2,a} \ge b) \\ &= E_{\chi^2_{2\nu}}(P(\chi^2_{2,a} + x \ge b|\chi^2_{2\nu} = x)) \\ &= \int_0^\infty f_{\chi^2_{2\nu}}(x)P(\chi^2_{2,a} \ge b-x)dx, \end{aligned} \quad (83)$$

where

$$P(\chi^2_{2,a} \ge b-x) = \begin{cases} Q_1\left(\sqrt{a}, \sqrt{b-x}\right), & 0 < x \le b, \\ 1, & x > b. \end{cases} \quad (84)$$

We first note that $x \mapsto P(\chi^2_{2,a} \geq x)$ is continuous on $(-\infty, \infty)$. Recently, we proved [44] that $b \mapsto Q_\nu(a, \sqrt{b})$ is strictly log-concave on $(0, \infty)$ when $a \geq 0$ and $\nu > 1$. Using the same proof process as in [44], we can also obtain that $b \mapsto Q_\nu(a, \sqrt{b})$ is log-concave on $(0, \infty)$ for $\nu = 1$. In view of part (e) of Lemma 5, we have that $x \mapsto Q_1(a, \sqrt{b-x})$ is log-concave on $(-\infty, b)$ for $a \geq 0$. Therefore, it is easy to prove that $x \mapsto P(\chi^2_{2,a} \geq b-x)$ is log-concave on $(-\infty, \infty)$, since $x \mapsto Q_1(a, \sqrt{b-x})$ is increasing and $x \mapsto P(\chi^2_{2,a} \geq b-x)$ is continuous on $(-\infty, \infty)$. Hence, $(x, y) \mapsto P(\chi^2_{2,a} \geq b-(x+y))$ is $SRR_2$ due to part (d) of Lemma 5. In view of (82) and the fact that $(x, \nu) \mapsto f_{\chi^2_{2\nu}}(x)$ is $TP_2$ for $x, \nu > 0$, we can use Lemma 4 to (83) and obtain that $(\nu_1, \nu_2) \mapsto Q_{\nu_1+\nu_2+1}\left(\sqrt{a}, \sqrt{b}\right)$ is $SRR_2$ for $\nu_1, \nu_2 > 0$. This implies that $\nu \mapsto Q_\nu\left(\sqrt{a}, \sqrt{b}\right)$ is log-concave for $\nu > 1$ and $a, b \geq 0$ due to part (d) of Lemma 5. Since $\nu \mapsto f_{\chi^2_{\nu,a}}$ is continuous for $\nu > 0$, the function $\nu \mapsto Q_\nu\left(\sqrt{a}, \sqrt{b}\right)$ is also continuous at $\nu = 1$ for $a, b \geq 0$. Hence, we can prove that $\nu \mapsto Q_\nu\left(\sqrt{a}, \sqrt{b}\right)$ is log-concave for $\nu \geq 1$. Substituting $a$ with $a^2$ and $b$ with $b^2$, the asserted result is proved.

(c) The function $Q_\nu(\sqrt{a}, \sqrt{b})$ can be also represented by using the non-central chi-square distribution with zero degrees of freedom, as follows

$$\begin{aligned}&Q_\nu\left(\sqrt{a}, \sqrt{b}\right)\\&= P(\chi^2_{2\nu,a} \geq b)\\&= P(\chi^2_{2\nu} + \chi^2_{0,a} \geq b)\\&= E_{\chi^2_{2\nu}}\left(P(x + \chi^2_{0,a} \geq b | \chi^2_{2\nu} = x)\right)\\&= \int_0^\infty f_{\chi^2_{2\nu}}(x) P(\chi^2_{0,a} \geq b-x) dx. \quad (85)\end{aligned}$$

All the same, $x \mapsto P(\chi^2_{0,a} \geq x)$ has a discontinuous point at $x = 0$, so it is not log-concave on the whole $(-\infty, \infty)$. However, by using (15) we get

$$\begin{aligned}&Q_\nu\left(\sqrt{a}, \sqrt{b}\right)\\&= \int_0^\infty f_{\chi^2_{2\nu}}(x) P(\chi^2_{0,a} \geq b-x) dx\\&= \int_b^\infty f_{\chi^2_{2\nu}}(x) dx + \int_0^b f_{\chi^2_{2\nu}}(x) P(\chi^2_{0,a} \geq b-x) dx\\&= Q_\nu\left(0, \sqrt{b}\right) + \int_0^b f_{\chi^2_{2\nu}}(x) P(\chi^2_{0,a} \geq b-x) dx, \quad (86)\end{aligned}$$

where

$$P\left(\chi^2_{0,a} \geq b\right) = \int_b^\infty \frac{1}{2}\left(\frac{a}{x}\right)^{1/2} e^{-\frac{x+a}{2}} I_1\left(\sqrt{ax}\right) dx \quad (87)$$

for all $b > 0$. Now, let us consider the normalized modified Bessel function of the first kind $\gamma_\nu : (0, \infty) \mapsto (1, \infty)$, defined by

$$\gamma_\nu(x) = 2^\nu \Gamma(\nu+1) x^{-\nu/2} I_\nu\left(\sqrt{x}\right). \quad (88)$$

We have

$$\begin{aligned}&P\left(\chi^2_{0,a} \geq b\right)\\&= \int_b^\infty \frac{1}{4} \gamma_1(ax) a e^{-\frac{x+a}{2}} dx\\&= \int_0^\infty \frac{1}{4} \gamma_1(ax) a e^{-\frac{x+a}{2}} \mathcal{L}_{[b,\infty)}(x) dx\\&= \int_0^\infty \frac{1}{4} \gamma_1(ax) a e^{-\frac{x+a}{2}} \mathcal{L}_{[0,\infty)}(x-b) dx. \quad (89)\end{aligned}$$

It is known that the function $x \mapsto \gamma_\nu(x)$ is log-concave if $\nu > -1$ [66, Theorem 2.2]. So $x \mapsto \gamma_1(ax)$ is also log-concave for $a > 0$. On the other hand, since the set $[0, \infty)$ is convex, the indicator function $x \mapsto \mathcal{L}_{[0,\infty)}(x)$ is log-concave. Hence, the function $(x, b) \mapsto \mathcal{L}_{[0,\infty)}(x-b)$ is also log-concave due to part (e) of Lemma 5. Therefore, the integrand of (89) is log-concave in $(x, b)$, as a product of log-concave functions.

It is known that if $(x, y) \mapsto f$ is log-concave, then $g(x) = \int f(x, y) dy$ is a log-concave function of $x$ [47, p. 106]. Applying this to (89), we obtain that $b \mapsto P\left(\chi^2_{0,a} \geq b\right)$ is log-concave on $(0, \infty)$. It follows that the function $x \mapsto P(\chi^2_{0,a} \geq b-x)$ is also log-concave on $(0, b)$, due to part (e) of Lemma 5. Then, by the same method used in part (a) and (b), we can prove that the function $\nu \mapsto \int_0^b f_{\chi^2_{2\nu}}(x) P(\chi^2_{0,a} \geq b-x) dx$ is log-concave on $(0, \infty)$. Hence, $\nu \mapsto Q_\nu(\sqrt{a}, \sqrt{b}) - Q_\nu(0, \sqrt{b})$ is log-concave on $(0, \infty)$. Substituting $a$ with $a^2$ and $b$ with $b^2$, the asserted result is proved.

(d) Similar to (85), we obtain that

$$\begin{aligned}&1 - Q_\nu\left(0, \sqrt{b}\right)\\&= P(\chi^2_{2\nu} \leq b)\\&= \int_0^\infty f_{\chi^2_{2\nu}}(x) \mathcal{L}_{(0,b)}(x) dx. \quad (90)\end{aligned}$$

where $\mathcal{L}_{(0,b)}$ is the indicator function for the interval $(0, b)$. We can easily prove that $(x, y) \mapsto \mathcal{L}_{(0,b)}(x+y)$ is $SRR_2$ for $x, y > 0$ and $b \geq 0$ from the definition. Thus, using the same method as in the proof of part (a) and (b), we can prove that the function $\nu \mapsto 1 - Q_\nu(0, b)$ is log-concave on $(0, \infty)$ for each $b > 0$.

(e) Similar to (85), we obtain that

$$\begin{aligned}&1 - Q_{\nu+1}\left(\sqrt{a}, \sqrt{b}\right)\\&= P(\chi^2_{2\nu+2,a} \leq b)\\&= P(\chi^2_{2\nu} + \chi^2_{2,a} \leq b)\\&= E_{\chi^2_{2\nu}}(P(\chi^2_{2,a} + x \leq b | \chi^2_{2\nu} = x))\\&= \int_0^b f_{\chi^2_{2\nu}}(x) P(\chi^2_{2,a} \leq b-x) dx\\&= \int_0^b f_{\chi^2_{2\nu}}(x) \left(1 - Q_1\left(\sqrt{a}, \sqrt{b-x}\right)\right) dx. \quad (91)\end{aligned}$$

We have proved the function $b \mapsto 1 - Q_1(a, \sqrt{b})$ is log-concave on $(0, \infty)$ for all $a \geq 0$ in Theorem 2. It follows that the function $x \mapsto 1 - Q_1(a, \sqrt{b-x})$ is log-concave on $(0, b)$, due to part (e) of Lemma 5. Thus, using the same method as in the proof of part (a) and (b), we can prove that the function $\nu \mapsto 1 - Q_\nu(a, b)$ is log-concave on $[1, \infty)$ for each $a \geq 0$, $b > 0$, and with this the proof of this theorem is complete.

# APPENDIX C
## PROOF OF LEMMA 6

We know that

$$\int_c^d f_{\chi^2_{2\nu,a_1+a_2}}(x)dx$$
$$= P\left(\chi^2_{2\nu,a_1+a_2} \in (c,d)\right)$$
$$= P\left(\chi^2_{2\nu,a_1} + \chi^2_{0,a_2} \in (c,d)\right)$$
$$= E_{\chi^2_{2\nu,a_1}}\left(P(x + \chi^2_{0,a_2} \in (c,d)|\chi^2_{2\nu,a_1} = x)\right)$$
$$= \int_0^d P\left(x + \chi^2_{0,a_2} \in (c,d)\right) f_{\chi^2_{2\nu,a_1}}(x)dx, \quad (92)$$

where we used that $P\left(x + \chi^2_{0,a_2} \in (c,d)\right) = 0$ for all $x > d$. Recall that [46, p. 437]

$$P\left(x + \chi^2_{0,a_2} \in (c,d)\right)$$
$$= \sum_{j=0}^\infty \frac{\left(\frac{1}{2}a_2\right)^j}{j!} e^{-a_2/2} P\left(x + \chi^2_{2j} \in (c,d)\right), \quad (93)$$

for all $0 < x < d$. Moreover, for all $j \geq 0$ integer $P(x+\chi^2_{2j} \in (c,d))$ can be further expressed as

$$P\left(x + \chi^2_{2j} \in (c,d)\right)$$
$$= \int_0^\infty f_{\chi^2_{2j}}(t) \mathcal{L}_{\{x+t \in (c,d), x, t > 0\}}(x,t)dt, \quad (94)$$

where $\mathcal{L}_{\{x+t \in (c,d), x, t > 0\}}(x,t)$ is the indicator function of the set $S = \{(x,t)|x + t \in (c,d), x, t > 0\}$.

Observe that $(x,t) \mapsto \mathcal{L}_{\{x+t \in (c,d), x, t > 0\}}$ is $SRR_2$ for $x, t > 0$ from the definition. Recall that from the proof of part (a) of Theorem 3 we already know that $(t,j) \mapsto f_{\chi^2_{2j}}(t)$ is $TP_2$ for $t > 0$ and $j \geq 0$ integer. Applying Lemma 3 for (94), we obtain that $(x,j) \mapsto P(x+\chi^2_{2j} \in (c,d))$ is $SRR_2$. From parts (a) and (b) of Lemma 5, we have that

$$(j, a_2) \mapsto \frac{\left(\frac{1}{2}a_2\right)^j}{j!} e^{-a_2/2} \quad (95)$$

is $TP_2$ for all $j \geq 0$ integer. Using again Lemma 3 for (93), we obtain that $(a_2, x) \mapsto P(x + \chi^2_{0,a_2} \in (c,d))$ is $SRR_2$ for $a_2 > 0$ and $0 < x < d$. It is known that $(x, a_1) \mapsto f_{\chi^2_{2\nu,a_1}}(x)$ is $TP_\infty$ on $x$ and $a_1$ with constant $\nu > 0$ [57, p. 110]. Then by another application of Lemma 3 for (92), we deduce that

$$(a_1, a_2) \mapsto \int_c^d f_{\chi^2_{2\nu,a_1+a_2}}(x)dx \quad (96)$$

is $SRR_2$ for $a_1$ and $a_2 > 0$. This implies that indeed the function $a \mapsto \int_c^d f_{\chi^2_{2\nu,a}}(x)dx$ is log-concave for $a > 0$ due to part (d) of Lemma 5. Since the pdf of non-central chi-square distribution given in (7) is replaced by the pdf of central chi-square distribution given in (10) when $a \to 0$, the function $a \mapsto \int_c^d f_{\chi^2_{2\nu,a}}(x)dx$ is continuous at $a = 0$. Hence it is also is log-concave for $a \geq 0$, which completes the proof.

# APPENDIX D
## PROOF OF THEOREM 5

(a) We still use the normalized modified Bessel function of the first kind $\gamma_\nu$, which is defined in (88), to represent the Nuttall $Q$-function. When $a > 0$, we have that

$$\mathcal{Q}_{\mu,\nu}\left(\sqrt{a}, \sqrt{b}\right)$$
$$= \int_b^\infty \frac{x^{(\mu-1)/2}}{2a^{\nu/2}} e^{-(x+a)/2} I_\nu\left(\sqrt{ax}\right)dx$$
$$= \int_b^\infty \gamma_\nu(ax) \frac{x^{(\mu+\nu-1)/2}}{2^{\nu+1}\Gamma(\nu+1)} e^{-(x+a)/2}dx$$
$$= \int_0^\infty \gamma_\nu(ax) \frac{x^{(\mu+\nu-1)/2}}{2^{\nu+1}\Gamma(\nu+1)} e^{-(x+a)/2} \mathcal{L}_{[b,\infty)}(x)dx$$
$$= \int_0^\infty \gamma_\nu(ax) \frac{x^{(\mu+\nu-1)/2}}{2^{\nu+1}\Gamma(\nu+1)} e^{-(x+a)/2} \mathcal{L}_{[0,\infty)}(x-b)dx. \quad (97)$$

Since the function $x \mapsto \gamma_\nu(x)$ is log-concave if $\nu > -1$ [66, Theorem 2.2], $x \mapsto \gamma_\nu(ax)$ is log-concave for $a > 0$. The function $x \mapsto \frac{x^{(\mu+\nu-1)/2}}{2^{\nu+1}\Gamma(\nu+1)} e^{-(x+a)/2}$ is also log-concave when $\mu + \nu \geq 1$. On the other hand, since the set $[0, \infty)$ is convex, the indicator function $x \mapsto \mathcal{L}_{[0,\infty)}(x)$ is log-concave. Hence, the function $(x, b) \mapsto \mathcal{L}_{[0,\infty)}(x - b)$ is also log-concave due to part (e) of Lemma 5. Therefore, the integrand of (97) is log-concave in $(x, b)$ as the product of several log-concave functions when $\mu + \nu \geq 1$ and $a > 0$. Recall that if $(x, y) \mapsto f$ is log-concave, then $g(x) = \int f(x,y)dy$ is a log-concave function of $x$ [47, p. 106]. Applying this to (97) and changing $a$ with $a^2$, part (a) is proved.

(b) Since $b \mapsto \mathcal{Q}_{\mu,\nu}(a, \sqrt{b})$ is log-concave and decreasing, the function $g(b) = b^2$ is convex, then the composite function $b \mapsto (f \circ g)(b) = \mathcal{Q}_{\mu,\nu}(a,b)$ is log-concave too [47, p. 84], and part (b) is proved.

(c) When $a > 0$, $E_{\chi_{2(\nu+1),\sqrt{a}}}(X^{\mu-\nu-1}) - \mathcal{Q}_{\mu,\nu}(\sqrt{a}, \sqrt{b})$ can be expressed as

$$E_{\chi_{2(\nu+1),\sqrt{a}}}(X^{\mu-\nu-1}) - \mathcal{Q}_{\mu,\nu}\left(\sqrt{a}, \sqrt{b}\right)$$
$$= \int_0^b \frac{x^{(\mu-1)/2}}{2a^{\nu/2}} e^{-(x+a)/2} I_\nu\left(\sqrt{ax}\right)dx$$
$$= \int_0^b \gamma_\nu(ax) \frac{x^{(\mu+\nu-1)/2}}{2^{\nu+1}\Gamma(\nu+1)} e^{-(x+a)/2}dx$$
$$= \int_0^\infty \gamma_\nu(ax) \frac{x^{(\mu+\nu-1)/2}}{2^{\nu+1}\Gamma(\nu+1)} e^{-(x+a)/2} \mathcal{L}_{(-\infty,b]}(x)dx$$
$$= \int_0^\infty \gamma_\nu(ax) \frac{x^{(\mu+\nu-1)/2}}{2^{\nu+1}\Gamma(\nu+1)} e^{-(x+a)/2} \mathcal{L}_{(-\infty,0]}(x-b)dx. \quad (98)$$

Therefore, using the same method as in the proof of part (a), we can prove easily the asserted result.

(d) From (2), we obtain that

$$E_{\chi_{2(\nu+1),a}}(X^{\mu-\nu-1}) - \mathcal{Q}_{\mu,\nu}(a,b)$$
$$= \frac{1}{a^\nu} \int_0^b t^\mu e^{-\frac{t^2+a^2}{2}} I_\nu(at)dt$$
$$= \frac{1}{a^\nu} \int_0^\infty t^\mu e^{-\frac{t^2+a^2}{2}} I_\nu(at) \mathcal{L}_{(-\infty,0]}(x-b)dt. \quad (99)$$

We have proved that function $x \mapsto xI_\nu(x)$ is log-concave for each $\nu \geq 1/2$ [44, Proposition 2.1]. Since $(x,b) \mapsto$

$\mathcal{L}_{(-\infty,0]}(x-b)$ is log-concave as proved in part (a), the integrand in (99) is also log-concave in $x$ and $b$ for all $\mu \geq 1$, $\nu \geq 1/2$. Hence, using the same method in the proof of part (a), the asserted result follows.

## APPENDIX E
## PROOF OF LEMMA 7

We know that

$$\int_c^d x^{(\mu-\nu-1)/2} f_{\chi^2_{2(\nu+1),a}}(x)dx$$

$$= \int_c^d x^{(\mu-\nu-1)/2} \left(\int_0^x f_{\chi^2_{2\nu}}(t) f_{\chi^2_{2,a}}(x-t)dt\right) dx$$

$$= \int_0^d f_{\chi^2_{2\nu}}(t) \left(\int_{\max\{c,t\}}^d x^{(\mu-\nu-1)/2} f_{\chi^2_{2,a}}(x-t)dx\right) dt$$

$$= \int_0^d f_{\chi^2_{2\nu}}(t)$$

$$\times \left(\int_{\max\{c-t,0\}}^{d-t} (x+t)^{(\mu-\nu-1)/2} f_{\chi^2_{2,a}}(x)dx\right) dt. \quad (100)$$

Since $f_{\chi^2_{2,a}}(x) = 0$ for $x < 0$, we have

$$\int_c^d x^{(\mu-\nu-1)/2} f_{\chi^2_{2(\nu+1),a}}(x)dx$$

$$= \int_0^d f_{\chi^2_{2\nu}}(t) \left(\int_{c-t}^{d-t} (x+t)^{(\mu-\nu-1)/2} f_{\chi^2_{2,a}}(x)dx\right) dt$$

$$= \int_0^d f_{\chi^2_{2\nu}}(t)$$

$$\times \left(\int_0^\infty (x+t)^{(\mu-\nu-1)/2} f_{\chi^2_{2,a}}(x) \mathcal{L}_{[c-t,d-t]}(x)dx\right) dt$$

$$= \int_0^d f_{\chi^2_{2\nu}}(t)$$

$$\times \left(\int_0^\infty (x+t)^{(\mu-\nu-1)/2} f_{\chi^2_{2,a}}(x) \mathcal{L}_{[c,d]}(x+t)dx\right) dt. \quad (101)$$

When $\mu - \nu$ is fixed and no smaller than 1, $x \mapsto x^{(\mu-\nu-1)/2}$ is log-concave. We also have that the function $x \mapsto \mathcal{L}_{[c,d]}(x)$ is log-concave, since $[c,d]$ is a convex set. Then, from part (e) of Lemma 5, we can conclude that $(x,t) \mapsto (x+t)^{(\mu-\nu-1)/2}$ and $(x,t) \mapsto \mathcal{L}_{[c,d]}(x+t)$ are both log-concave. We know that the function $x \mapsto f_{\chi^2_{2,a}}(x)$ is log-concave on $(0,\infty)$ for all $a > 0$ [55]. Therefore, $(x,t) \mapsto (x+t)^{(\mu-\nu-1)/2} f_{\chi^2_{2,a}}(x) \mathcal{L}_{[c,d]}(x+t)$ is log-concave as a product of log-concave functions. Hence, we have that $g(t) = \int_0^\infty (x+t)^{(\mu-\nu-1)/2} f_{\chi^2_{2,a}}(x) \mathcal{L}_{[c,d]}(x+t)dx$ is log-concave in $t$, and thus $(t_1,t_2) \mapsto g(t_1+t_2)$ is $SRR_2$ due to part (d) of Lemma 5.

On the other hand it is known that

$$f_{\chi^2_{2(\nu_1+\nu_2)}}(x) = f_{\chi^2_{2\nu_1}}(x) * f_{\chi^2_{2\nu_2}}(x)$$

$$= \int_0^x f_{\chi^2_{2\nu_1}}(t) f_{\chi^2_{2\nu_2}}(x-t)dt, \quad (102)$$

and $(x,\nu) \mapsto f_{\chi^2_{2\nu}}(x)$ is $TP_2$ for $x, \nu > 0$. Then, applying Lemma 4 in (101), we obtain that $(\nu_1,\nu_2) \mapsto$ $\int_c^d x^{(\mu-\nu_1-\nu_2-1)/2} f_{\chi^2_{2(\nu_1+\nu_2+1),a}}(x)dx$ is $SRR_2$ for $\nu_1, \nu_2 > 0$, which implies that $\nu \mapsto \int_c^d x^{(\mu-\nu-1)/2} f_{\chi^2_{2(\nu+1),a}}(x)dx$ is log-concave for $\nu > 0$ and fixed $\mu - \nu \geq 1$, due to part (d) of Lemma 5. Moreover, it can be easily verified that $\nu \mapsto \int_c^d x^{(\mu-\nu-1)/2} f_{\chi^2_{2(\nu+1),a}}(x)dx$ is continuous at $\nu = 0$. Therefore, $\nu \mapsto \int_c^d x^{(\mu-\nu-1)/2} f_{\chi^2_{2(\nu+1),a}}(x)dx$ is log-concave on $[0,\infty)$ for fixed $\mu - \nu \geq 1$, and the proof is complete.

## APPENDIX F
## PROOF OF LEMMA 8

Similarly with the argument of Lemma 7, we can get

$$\int_c^d x^{(\mu-\nu-1)/2} f_{\chi^2_{2(\nu+1),a_1+a_2}}(x)dx$$

$$= \int_0^d f_{\chi^2_{2\nu,a_1}}(t)$$

$$\times \left(\int_0^\infty f_{\chi^2_{2,a_2}}(x)(x+t)^{(\mu-\nu-1)/2} \mathcal{L}_{[c,d]}(x+t)dx\right) dt. \quad (103)$$

When $\mu - \nu$ is fixed and no smaller than 1, $x \mapsto x^{(\mu-\nu-1)/2}$ is log-concave. We also have that the function $x \mapsto \mathcal{L}_{[c,d]}(x)$ is log-concave, since $[c,d]$ is a convex set. Then, the function $x \mapsto x^{(\mu-\nu-1)/2} \mathcal{L}_{[c,d]}(x)$ is also log-concave as a product of two log-concave functions. From part (d) of Lemma 5, we can conclude that the function $(x,t) \mapsto (x+t)^{(\mu-\nu-1)/2} \mathcal{L}_{[c,d]}(x+t)$ is $SRR_2$ for $x,t > 0$. It is known that $(x,a_2) \mapsto f_{\chi^2_{2,a_2}}(x)$ is $TP_\infty$ on $x$ and $a_2$ [57, p. 110]. Using Lemma 3, we obtain that the function

$$g(t,a_2) = \int_0^\infty f_{\chi^2_{2,a_2}}(x)(x+t)^{(\mu-\nu-1)/2} \mathcal{L}_{[c,d]}(x+t)dx \quad (104)$$

is $SRR_2$ for $t, a_2 > 0$, and

$$\int_c^d x^{(\mu-\nu-1)/2} f_{\chi^2_{2(\nu+1),a_1+a_2}}(x)dx$$

$$= \int_0^d f_{\chi^2_{2\nu,a_1}}(t) g(t,a_2)dt. \quad (105)$$

We know that $(t,a_1) \mapsto f_{\chi^2_{2\nu,a_1}}(t)$ is also $TP_\infty$ for $t, a_1 > 0$ when $\nu > 0$. Therefore, by another application of Lemma 3 in (105), we get that $(a_1,a_2) \mapsto \int_c^d x^{(\mu-\nu-1)/2} f_{\chi^2_{2(\nu+1),a_1+a_2}}(x)dx$ is $SRR_2$ for $\nu, a_1, a_2 > 0$. This implies that the function $a \mapsto \int_c^d x^{(\mu-\nu-1)/2} f_{\chi^2_{2(\nu+1),a}}(x)dx$ is log-concave on $(0,\infty)$ for $\nu > 0$ and $\mu - \nu \geq 1$ fixed, due to part (d) of Lemma 5. Since $\nu \mapsto \int_c^d x^{(\mu-\nu-1)/2} f_{\chi^2_{2(\nu+1),a}}(x)dx$ is continuous at $\nu = 0$, we can prove that the log-concavity holds true for the case $\nu = 0$. Thus, the proof is complete.

**Yin Sun** (S'08) received the B. Eng. degree in electronic engineering from Tsinghua University, Beijing, China, in 2006. He is currently pursuing the Ph.D. degree at the Department of Electronic Engineering, Tsinghua University, Beijing, China. His research interests include optimization, information theory and wireless communication. He is now working on channel fading statistics and radio resource management of cognitive radio and cooperative communication systems.

**Árpád Baricz** received his B.Sc., M.Sc. and Ph.D. degrees in mathematics from the Babeş-Bolyai University, Cluj-Napoca, Romania, in 2003, 2004 and 2008, respectively. He received also a Ph.D. degree in mathematics from the University of Debrecen, Institute of Mathematics, Hungary, in 2008. Currently, he is a Senior Lecturer at Babeş-Bolyai University, Department of Economics, Romania. His research interests include special functions, geometric function theory, classical analysis and probability theory.

**Shidong Zhou** (M'98) is a professor at Tsinghua University, China. He received Ph.D. degree in communication and information systems from Tsinghua University in 1998. His B.S. and M.S. degrees in wireless communications were received from Southeast University, Nanjing, in 1991 and 1994, respectively. From 1999 to 2001 he was in charge of several projects in the China 3G Mobile Communication R&D Project. He is now a member of the China FuTURE Project, targeting on B3G technique and systems. His research interests are in the area of wireless and mobile communications.